\documentclass[twocolumn,showpacs,preprintnumbers,prb]{revtex4-1}

\usepackage{epsfig}
\usepackage{graphicx}
\usepackage{dcolumn}
\usepackage{bm}
\usepackage{color}
\usepackage{amstext}

\newcommand{\bra}[1]{\langle{#1}\vert}
\newcommand{\ket}[1]{\vert{#1}\rangle}
\newcommand{\bracket}[1]{\langle{#1}\rangle}
\newcommand{\BesselJ}[0]{\mathcal{J}}
\newcommand{\StruveH}[0]{\mathcal{H}}

\begin{document}

\title{Numerical method for non-linear steady-state transport in one-dimensional correlated conductors}

\author{M. Einhellinger}
\affiliation{Institut f\"ur Theoretische Physik, Leibniz Universit\"at Hannover, Appelstrasse 2, D-30167 Hannover, Germany}

\author{A. Cojuhovschi}
\affiliation{Institut f\"ur Theoretische Physik, Leibniz Universit\"at Hannover, Appelstrasse 2, D-30167 Hannover, Germany}

\author{E. Jeckelmann}
\affiliation{Institut f\"ur Theoretische Physik, Leibniz Universit\"at Hannover, Appelstrasse 2, D-30167 Hannover, Germany}

\date{\today}

\begin{abstract}
	We present a method for investigating the steady-state transport properties of
	one-dimensional correlated quantum systems. Using a procedure based on our
	analysis of finite-size effects in a related classical model ($LC$ line)
	we show that stationary currents can be obtained from transient currents
	in finite systems driven out of equilibrium. The non-equilibrium 
	dynamics of correlated quantum systems is calculated using the 
	time-evolving block decimation method. To demonstrate our method we determine
	the full $I$--$V$ characteristic of the spinless fermion model with 
	nearest-neighbour
	hopping $t_{\text{H}}$ and interaction $V_{\text{H}}$ using two different setups to generate 
	currents (turning on/off a potential bias). Our numerical results agree with 
	exact results for non-interacting fermions ($V_{\text{H}}=0$).
	For interacting fermions we find that in the linear regime $eV \ll 4t_{\text{H}}$ the 
	current $I$ is independent from the setup and our numerical data agree with the 
	predictions of the Luttinger liquid theory combined with the Bethe Ansatz 
	solution. For larger potentials $V$ the steady-state current depends on 
	the current-generating setup and as $V$ increases we find a negative 
	differential conductance with one setup while the currents saturate at finite 
	values in the other one. Both effects are due to finite renormalized bandwidths.
\end{abstract}


\maketitle

\section{Introduction}

	Much of our understanding of electronic transport in solids is based 
	on a picture of weakly-interacting charge carriers such as Fermi 
	liquid quasi-particles.	One-dimensional electron systems are well-known examples
	where this approximation fails. 
	The low-energy physics of these systems is described by the Tomonaga-Luttinger
	liquid (TLL) theory and is realized in quantum wires such as carbon nanotubes, 
	nanowires in semiconductor heterostructures, or metal atomic chains on surface 
	substrates.\cite{Schoenhammer02,Giamarchi03,Schoenhammer04}
	
	The transport properties of one-dimensional correlated systems 
	have been extensively studied during the last two 
	decades.\cite{Giamarchi03,Schoenhammer04,Yacoby04,Zotos04,Kawabata07}
	A major goal is to determine and understand the current-voltage characteristic $I-V$ 
	of various systems made of quantum dots and wires.
	Most of these studies have been restricted to a regime where
	the energy scale of the current-generating electromagnetic field or potential 
	bias is small in comparison to the energy scale of the unperturbated
	systems (i.e., the band width in lattice models and the Fermi velocity in
	continuum models). The non-linear regime has mostly been investigated in
	quantum contact problems, where the interaction is confined to a small region
	of the system. 
	For instance, non-linear current-voltage characteristics have been calculated within the
	TLL theory such as the power-law $I \sim V^{\alpha}$ behaviour for the transport
	through a weak link \cite{Kane92,Furusaki93} or  the exact $I(V)$ curve for the
	current through a point contact in a fractional
	quantum Hall edge state device.\cite{Fendley95,Komnik11}
	However, the TLL theory is limited to low-energy excitations with linear dispersion
	and thus to potential biases $V$ which are weak compared to the band width. 
	Only recently the implications of a non-linear dispersion have started to be 
	considered.\cite{Barak10}
	Thus current-voltage characteristics from the TLL theory 
	are actually limited to a weak-bias regime.
	
	There are few works presenting full current-voltage characteristics with a voltage 
	up to the largest energy scale of the system (for instance,
    see~Refs.~\onlinecite{Boulat08}, ~\onlinecite{Heidrich09a}, and~\onlinecite{Carr11})
	and none is concerned with one-dimensional correlated conductors.
	Thus transport properties beyond linear response are poorly understood.
	We believe that it is important to attain a better knowledge of the non-linear
	transport properties in one-dimensional correlated conductors. First,
	the validity of weak-bias approaches can be confirmed only if
	one obtains some quantitative estimates of non-linear effects. Moreover, 
	non-linear devices play a significant role in electronics and 
	studies of non-linear dynamics are required to reveal the full potential
	functionality of quantum wires as electronic circuit components.
	
	In this work we develop and apply a method
	for investigating the zero-temperature DC transport properties of one-dimensional
	correlated conductors for potential biases up to the order of the band width.
	For this purpose we study a well-known one-dimensional lattice model described by 
	the half-filled spinless fermion Hamiltonian with nearest-neighbour repulsion.
	Even though this model is exactly solvable by the Bethe Ansatz and
	the low-energy physics are determined by the generic TLL phenomenology\cite{Giamarchi03,Schoenhammer04},
	its transport properties in the non-linear regime cannot be obtained analytically.
	
	To determine the transport properties we simulate the quantum dynamics of single chains 
	which are driven out of equilibrium by a potential bias between the left- and right-hand halves of the chain
	and calculate the resulting currents through the middle bond.
	The simulation of out-of-equilibrium quantum many-particle systems
	is one of the major challenges in computational physics. Recently,
	a family of numerical methods has been developed to simulate
	the real-time evolution of quantum lattice systems such as
	one-dimensional Hamiltonians with short-range interactions.\cite{Schollwoeck06,Noack08,Schollwoeck11}
	The most prominent ones are the time-dependent Density-Matrix Renormalization Group (td-DMRG)
	and the time-evolving block decimation (TEBD) method.\cite{Vidal03,Vidal04}
	Various flavours of td-DMRG have been successfully applied in studies of quantum 
	many-body dynamics. In particular, they have proven to be promising tools
	for investigating electronic transport in strongly correlated nanostructures and one-dimensional    conductors.\cite{Boulat08,Heidrich09a,Cazalilla02,Luo03,Cazalilla03,Schmitteckert04,Al-Hassanieh06,Kirino08,Heidrich09b,Kirino10,Heidrich10,Znidaric11,Jesenko11}
	
	Surprisingly, the original TEBD method\cite{Vidal03,Vidal04} has not been applied to electronic transport problems yet.
	Both td-DMRG and TEBD methods can be described within a common mathematical framework, 
	the matrix product quantum states.\cite{Schollwoeck11,Daley04,Jeckelmann08}
	Their accuracy and efficiency depend essentially on 
	the amount of entanglement in the quantum system and should be similar.
	Admittedly, the TEBD algorithm is restricted to a small family of systems (one-dimensional
	Hamiltonians or ladder systems with nearest-neighbour interactions only) while td-DMRG techniques are more versatile.
	However, the TEBD algorithm is naturally parallelizable and thus fully scalable, which is a required
	feature in high-performance computing, while the efficient parallelization of the DMRG algorithms
	remains an open challenge.\cite{Hager04,Schmitteckert07}
	In this work we employ the TEBD method for computing the non-equilibrium
	dynamics of the spinless fermion model.
	
	With the TEBD method we can compute the non equilibrium quantum dynamics of 
	lattice models with a finite number of sites $N$ over a finite period of time $t$.  
	As we are primarily interested in determining the DC transport properties,
	TEBD results must be extrapolated to the thermodynamic limit $N\rightarrow \infty$
	and to the steady-state limit $t \rightarrow \infty$. However, finite-size and finite-time
	effects are very complex in these out-of-equilibrium quantum systems (for 
	instance, see~Ref.~\onlinecite{Branschaedel10})
	and extrapolations are difficult because we know very little about 
	the scaling of currents with system size $N$ and time $t$.
	For this reason we have investigated this scaling in an exactly solvable classical model, the so-called $LC$ line.
	Using this information we have developed a method for quantum systems which allows us
	to obtain reliable quantitative results for stationary currents $I$ from numerical data
	for rather small system sizes and short simulation times.
	
	In this paper we show that our extrapolation approach allows one to determine
	the full $I-V$ curves of interacting one-dimensional conductors using the 
	spinless fermion model for illustration.
	In the non-interacting case this model can be solved exactly using the equation of motion method.
	The outcomes of this special case confirm our extrapolation method and reveal 
	a negligible numerical error for the TEBD simulation results.
	For interacting fermions comparisons with predictions from the TLL theory and the Bethe-Ansatz 
	solution confirm the validity of our  method in the linear response regime.
	While for the linear regime the specific setup does not matter,
	it is highly decisive for the non-linear current-voltage characteristics.
	We basically distinguish between two different ways of creating a current flow.\cite{Branschaedel10}
	In the first setup (I) we apply an initial voltage and calculate the ground state
	which has different particle numbers in its two halves
	and then let the system evolve with an overall equal on-site potential.
	In the second setup (II) we calculate the ground state without a potential difference but turn it on for the real time evolution.
	We have found that while the general shape of the current-curve as a function of time is dominated by classical effects
	(i.e., which are also found in the $LC$ line model), 
	the current-voltage characteristics are primarily determined by the chosen setup.
	For setup (I) the system shows a positive differential conductance for the full voltage-range and saturates at a finite value
	for very large potential differences.
	For the second setup (II) the linear response  coincides with the one for setup (I)
	but for higher voltages  we observe a negative differential conductance.
	Both effects are also present in the non-interacting case and come from the finite bandwidth 
	and the non-linear dispersion (i.e., the energy-dependent density of states) of the 
	excitations.\cite{Branschaedel10,Baldea10}
	
	Our paper is organized as follows: In the next section we introduce the spinless fermion model,
	and the methods used in this work.
	In the third section we investigate the classical $LC$ line to explain the basic behaviour of the current in
	finite and infinite one-dimensional chains and derive our method to extrapolate the stationary current
	from finite system results.
	In the fourth section we describe finite-size effects and the convergence to a stationary current for infinite system sizes,
	while the $I-V$ characteristic of the spinless fermion model and comparisons with exact results are shown in the fifth section.
	Finally, we summarize our findings in the last section.
    Some calculations are detailed in appendices.

\section{Model and methods}

	\subsection{Spinless fermion model}
		
		We consider a one-dimensional lattice model representing correlated conductors driven out of 
		equilibrium by a potential bias. For spinless fermions the Hamiltonian without a potential bias is
		\begin{eqnarray}\label{eq:spinlessHamiltonian}
			H_0 = &&- t_{\text{H}} \sum_{j=1}^{N-1} (c_{j}^{\dagger}c_{j+1}^{} +c_{j+1}^{\dagger}c_{j}^{})\nonumber\\
			      &&+ V_{\text{H}} \sum_{j=1}^{N-1}\left (n_{j}-\frac{1}{2} \right ) \left (n_{j+1}	- \frac{1}{2} \right )
		\end{eqnarray}
		where $t_{\text{H}}$ denotes the hopping amplitude between nearest-neighbour sites, $V_{\text{H}}$ is the Coulomb repulsion between spinless fermions on 
		nearest-neighbour sites,
		and $n_{j} = c_{j}^{\dagger}c_{j}^{}$. At half filling ($N/2$ fermions in the $N$-site lattice)
		this Hamiltonian describes an ideal conductor for $-2t_{\text{H}} < V_{\text{H}} \leq 2t_{\text{H}}$. 
		It can be interpreted as a system of spin-polarized electrons.
		
		The low-energy behaviour of this lattice model is described by the TLL 
		theory.\cite{Giamarchi03,Schoenhammer04} The generic properties of a TLL are determined 
		by two parameters: The velocity of elementary excitations (renormalized Fermi velocity) $v$
		and a dimensionless parameter $K$.
		From the Bethe Ansatz solution we know the relation between TLL parameters and the parameters 
		$t_{\text{H}}$ and $V_{\text{H}}$ of the spinless fermion model at half filling
		\begin{equation}\label{eq:velocity}
			v = \pi \frac{a_L t_{\text{H}}}{\hbar} \sqrt{1-\left
            (\frac{V_{\text{H}}}{2t_{\text{H}}} \right )^2} \left [ \arccos
			\left (\frac{V_{\text{H}}}{2t_{\text{H}}} \right ) \right ]^{-1}
		\end{equation}
		and
		\begin{equation}\label{eq:TLLparameter}
			K = \frac{\pi}{2} \left[\pi - {\arccos}\left ( \frac{V_{\text{H}}}{2t_{\text{H}}} \right )\right]^{-1}
		\end{equation}
		where $a_L$ is the lattice constant. 
		To drive the system out of equilibrium we a use step-like potential bias between the
		left- and right-hand halves of the chain
		\begin{equation}\label{eq:H_B_def}
			H_B = \frac{\Delta\epsilon}{2} \left( \sum_{j=1}^{N/2} n_j - \sum_{j=N/2+1}^{N} n_j \right) .
		\end{equation}
		The potential energy step is set by $\Delta\epsilon = |eV|$ where $V$ is the voltage bias
		and $e$ is the elementary charge.
		It is possible to use a smoother potential profile but the results for the stationary
		current using our extrapolation method are only slightly affected by the specific shape as long as the non-constant
		part in the middle is rather smooth and locally confined.

		\begin{figure}[t]
			\begin{center}
				\includegraphics[width=0.95\hsize]{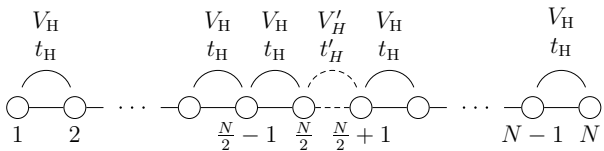}
				\caption{One-dimensional conductor consisting of two coupled leads.
				 				 A bias between its right- and left-hand halves is applied and the current is measured in the middle
				         of the system at the junction between both sides (dashed connection).
				\label{fig:setup}}
			\end{center}
		\end{figure}
		
		The above system can be seen as two coupled interacting leads made of the sites 
		$\{ 1 \dots \frac{N}{2}\}$ and $\{\frac{N}{2}+1 \dots N\}$, 
		respectively, see Fig.~\ref{fig:setup}. The coupling is given by a hopping term $t'_H$ between 
		the left- and right-hand sides of the system (i.e., between site $N/2$ and site $N/2+1$)
		and an additional coupling $V'_{\text{H}} (\ n_{N/2}-1/2) (\ n_{N/2+1}-1/2)$.
		Here we discuss the homogeneous system only ($t'_H=t_{\text{H}}$ and $V'_{\text{H}} = V_{\text{H}}$)
		but our approach can be easily extended to systems with a weak link $t'_H  < t_{\text{H}}$ representing the 
		tunneling through a nanostructure\cite{Cazalilla02} or with a site representing a quantum dot,
		such as the Interacting Resonant Level Model (IRLM)\cite{Boulat08}, 
		as well as to systems with a few additional sites intercalated between both leads and
		representing nanostructures with internal degrees of freedom.\cite{Schmitteckert04}

		The current operator between the pair of sites $(k,k+1)$ is 
		\begin{equation}\label{eq:current_op}
			j_{k} = i\frac{e t_{\text{H}}}{\hbar} \left ( c_{k}^{\dagger} c_{k+1}^{} - c_{k+1}^{\dagger} c_{k}^{} \right ) .
		\end{equation}
		For a given (time-dependent) quantum state we define the current flowing
		between both halves of the system (see Fig.~\ref{fig:setup}) as the expectation value of the current
		operator for the site pair in the middle of the system
		\begin{equation}\label{eq:current}
			J(t) = \left \langle j_{N/2} \right \rangle .
		\end{equation}
			We note that 
		\begin{equation}
			J(t) = - \frac{d}{dt}Q_L(t) = \frac{d}{dt}Q_R(t)
		\end{equation}
		where 
		\begin{equation}
			Q_L(t) = -e \sum_{k=1}^{N/2}  \langle n_{k} \rangle
			\quad \text{and} \quad
			Q_R(t) = -e \sum_{k=N/2+1}^{N} \langle n_{k} \rangle 
		\end{equation}
		are the (time-dependent) charges in the left- and right-hand halves of the chain, respectively.
		As the number of particles $n$ is conserved in our models, $Q_L(t) + Q_R(t) = -e\cdot n = \text{const}$.
		The stationary current is a constantly flowing current in an infinitely large system after the settling time
		\begin{equation}
			\bar{J} = \lim_{t\rightarrow\infty} \lim_{N\rightarrow\infty} J(t)
		\end{equation}
		and according to the TLL theory in the linear regime (small $\Delta\epsilon$) it is given by\cite{Giamarchi03}
		\begin{equation}\label{eq:J_bar_sf}
				\bar{J} = \frac{e^2}{h}K V = \frac{e}{h} K \Delta\epsilon \quad \text{for} \quad V \ge 0.
		\end{equation}
		We set $t_{\text{H}}=a_L=1$ for all numerical simulations and, if units are not given explicitly, $e=\hbar=1$.
		
In this work both setups used to generate a current (see the next section) lead to an overall half-filled 
system. Thus in the weak potential bias regime both system halves remain approximately
 half-filled at all times.  Consequently, we expect that the current is given by the Luttinger
 liquid prediction 
 (\ref{eq:J_bar_sf}) together with the Bethe Ansatz result (\ref{eq:TLLparameter}) in the linear response regime.
		
		\subsection{Setups for non-equilibrium simulations}
		
		We employ two different setups to generate currents in the lattice 
		models. \cite{Branschaedel10}
		\begin{figure}[t]
			\begin{center}
				\includegraphics[width=1\hsize]{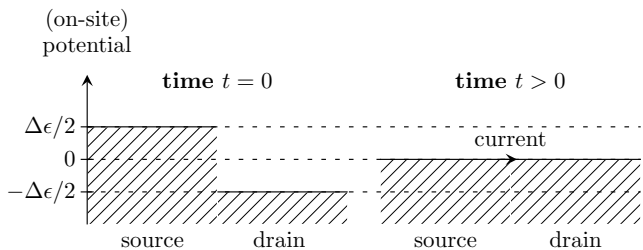}
				\caption{Setup (I): The charge reservoirs (two halves: source and drain) have different potentials but are coupled for $t=0$ while for $t>0$
				         the potential difference is set to $\Delta\epsilon = 0$ instantaneously.
				\label{fig:setup1}}
			\end{center}
		\end{figure}
		In the first one (I) we prepare the system at time $t=0$ in the ground state 
		$| \phi (\Delta \epsilon \neq 0) \rangle$ of the Hamiltonian
		$H = H_0 + H_B$ (i.e. with potential bias), see Fig.~\ref{fig:setup1}. For later times $t > 0$ we let the system evolve
		according to the Hamiltonian $H_0$ (i.e. without potential bias)
		\begin{equation}\label{eq:psi1}
			| \psi(t > 0) \rangle = \exp\left (-i\frac{H_0t}{\hbar}\right ) | \phi (\Delta \epsilon \neq 0) \rangle .
		\end{equation}
		This setup describes an inhomogeneous initial state with more particles in one half of the
		system than in the other one. Thus particles flow from one side to the other one for $t > 0$.
		It corresponds to a one-dimensional scattering experiment in which particles are 
		emitted on one side of the system with energies between 
		$[-\Delta\epsilon/2, \Delta\epsilon/2]$, scattered at the junction between both system halves,
		and then (partially) transmitted to the opposite side. This picture of transport through junctions is 
		often used in theoretical investigations.

		\begin{figure}[b]
			\begin{center}
				\includegraphics[width=1\hsize]{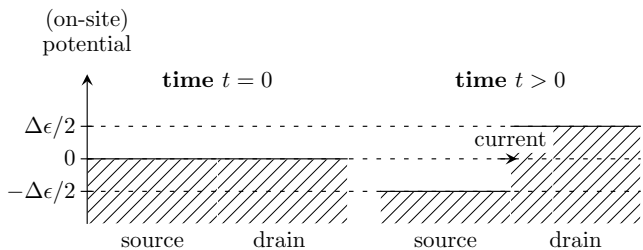}
				\caption{Setup (II): The charge reservoirs (two halves: source and drain) are in equilibrium and coupled with $t_{\text{H}}$ and $V_{\text{H}}$ for $t=0$.
	         			 For $t>0$ a potential difference $\Delta\epsilon > 0$ is applied.
				\label{fig:setup2}}
			\end{center}
		\end{figure}

		In the second setup (II) we prepare the system at time $t=0$ in the ground state 
		$| \phi (\Delta \epsilon = 0) \rangle$ of the Hamiltonian
		$H_0$ (i.e., without potential bias). For later times $t > 0$ the time evolution of the system is
		determined by the Hamiltonian $H=H_0 - H_B$, i.e. with a potential bias that causes the current to flow in the same direction as in setup (I)
		\begin{equation}\label{eq:psi2}
			| \psi(t > 0) \rangle = \exp\left (-i\frac{(H_0 - H_B)t}{\hbar}\right ) | \phi (\Delta \epsilon = 0) \rangle .
		\end{equation}
		Setup (II) describes the evolution of an initially homogeneous state under the influence
		of a potential gradient, see Fig.~\ref{fig:setup2}. 
		Thus it corresponds more closely to the actual experimental situation
		with a voltage source generating a current in a conducting wire.
		
		\noindent We should point out that in setup (I) the two leads are decoupled with respect to the Coulomb-interaction $V'_{\text{H}}$
		for the calculation of the ground state.
		Our tests have revealed that otherwise a strong dependency of the stationary current on the system size appears, that is,
		for smaller system sizes the stationary current becomes higher but for $N\rightarrow \infty$ it approaches the same constant value
		as the value one gets with $V'_{\text{H}}=0$ for $t=0$. Choosing $V'_{\text{H}}=0$ for the computation of the ground state 
		therefore decreases the finite-size error.

		Generally,  currents~(\ref{eq:current}) calculated with the states~(\ref{eq:psi1})
		and~(\ref{eq:psi2}) are different. In the strong-bias limit $|\Delta \epsilon| \gg
		 t_{\text{H}}, V_{\text{H}}$  it is easy to show that the steady-state current remains finite
		for the first setup while it vanishes for the second one.
		Recently, it has been reported that initial conditions (quenching an interaction term
		or a tunneling term) can also alter
		the steady-state current flowing through a quantum point contact 
		between two TLL leads which have been driven out of equilibrium by an external 
		bias.\cite{Perfetto10}
		In the weak-bias limit $|\Delta \epsilon| \ll t_{\text{H}}$, however, a simple perturbation
		calculation shows that both setups yield the same linear response for the stationary
		current. Thus in this regime both setups can be used indifferently but in the non-linear
		regime we must distinguish them.
		
		In most theoretical studies the potential bias switching is not instantaneous 
		but adiabatic. This can also be used in numerical simulations, for instance,
		see Refs.~\onlinecite{Cazalilla02} and~\onlinecite{Kirino10}. Our tests do not reveal
		any significant differences for the steady-state current depending
		on the switching rate as long as the potential changes in a time scale
        which is much smaller than
		the time scale associated with the motion of particles from one
        reservoir to the other one.
		Since our extrapolation method is designed to work with an instantaneous change in the potential difference
		and as numerical simulations are simpler with it, we prefer this approach.

	\subsection{One-particle equation of motion}
		
		Without Coulomb-interaction $V_{\text{H}}(=V'_{\text{H}})=0$ the spinless fermion model, described by the Hamiltonian (\ref{eq:spinlessHamiltonian}),
		reduces to the tight-binding model (non-interacting fermions without spin degree of freedom).
		A one-dimensional chain in the tight-binding model can be described by the single particle reduced density matrix
		\begin{equation}\label{eq:one_body_definition}
				\mathcal{G}_{ij}(t) := \bra{\Psi(t)}c_{i}^{\dagger}c_{j}^{}\ket{\Psi(t)}
		\end{equation}
		where the time evolution is given by the one-particle equation of motion
		\begin{equation}\label{eq:time_evolution_Greens_function}
				\frac{d}{dt} \mathcal{G}(t) = \frac{i}{\hbar} [H_{t>0}^{(1)}, \mathcal{G}(t)] .
		\end{equation}
		$H_{t>0}^{(1)}$ denotes the single particle Hamilton matrix of size $N\times N$ with which the system is evolved in time.
		More precisely $H_{t>0}^{(1)}$ is the one-particle representation of $H_0$ or $H_0+H_B$ depending on the specific setup.
		The particle number expectation values coincide with the diagonal terms of the reduced density matrix
		\begin{equation}
				\bracket{n_k(t)} = \bra{\Psi(t)} c_{k}^{\dagger}c_{k}^{} \ket{\Psi(t)} = \mathcal{G}_{kk}(t)
		\end{equation}
		and the expectation value for the current operator (\ref{eq:current_op})  from site $k$ to $k+1$  can be taken from off-diagonal entries
		\begin{equation}
				\bracket{j_k(t)} = i \frac{e t_{\text{H}}}{\hbar}  [\mathcal{G}_{k,k+1}(t) - \mathcal{G}_{k+1,k}(t)] .
		\end{equation}
		Consequently, the dynamics of a tight-binding chain can be computed numerically without additional truncation error
		with a runtime of $\mathcal{O}(N^3)$ for any point in time.

	\subsection{\label{tebd_method}TEBD method}
		
		The TEBD method\cite{Vidal03,Vidal04} is based on a specific representation of quantum states by matrix product states (MPS).
		For an $N$-site lattice it is generally written 
		\begin{eqnarray}\label{eq:mps_definition}
			| \Psi \rangle = &&\sum_{\{j_k\}} \Gamma^{[1]j_1}\lambda^{[1]} \Gamma^{[2]j_2}\lambda^{[2]} \dots \nonumber\\
											 &&\dots \Gamma^{[N-1]j_{N-1}}\lambda^{[N-1]} \Gamma^{[N]j_N} | j_1 j_2 \dots j_N \rangle
		\end{eqnarray}
		where $| j_1 j_2 \dots j_N \rangle$ designs the states of the occupation number basis,
		$\lambda^{[k]} (k=1,2,\dots,N-1)$ are positive definite diagonal matrices and 
		$\Gamma^{[k]j_k} (k=1,2,\dots,N)$ are matrices satisfying orthogonality conditions 
		\begin{eqnarray}
			\sum_{j_k} \left ( \Gamma^{[k]j_k} \right )^{\dagger} \left
            (\lambda^{[k-1]} \right )^2 \Gamma^{[k]j_k} &=& I ,\nonumber\\
			\sum_{j_k} \Gamma^{[k]j_k} \left ( \lambda^{[k]} \right )^2 \left ( \Gamma^{[k]j_k} \right)^{\dagger} &=& I .
		\end{eqnarray}
		Sums over an index $j_k$ run over a complete basis of the site $k$ (for instance 2 states for the spinless fermion model). 
		Every quantum state of the Fock space associated with a finite lattice can be represented
		exactly in this form if the matrix dimensions can be as large as the square root of the
		Fock space dimension (for instance $2^{N/2}$ for the spinless fermion model).
		In numerical computations, however, the matrix dimension must be kept smaller than
		a relatively small upper limit $\chi_c$. Fortunately, for many one-dimensional systems this truncation is possible and can lead to
		a dramatic computational speedup while keeping the error in computed observables conveniently low. This is done by using
		the Schmidt decomposition of the density matrix at each bond to calculate the matrices $\lambda^{[k]}$ from
		its eigenvalues and the $\Gamma^{[k]j_k}$ from its eigenvectors. Given a bond $k$ in the chain, the Schmidt decomposition at this
		bipartite split is defined by
		\begin{equation}
			\ket{\Psi} = \sum_{\alpha_{k}=1}^{\chi_k} \lambda_{\alpha_k} | \Phi_{\alpha_k}^{[1..k]} \rangle | \Phi_{\alpha_k}^{[k+1..N]} \rangle .
		\end{equation}
		The vectors $| \Phi_{\alpha_k}^{[1..k]} \rangle$ are the eigenvectors of $\rho^{[1..k]}$, the reduced density matrix of the left side of the split, 
		the $| \Phi_{\alpha_k}^{[k+1..N]} \rangle$ correspondingly the eigenvectors of $\rho^{[k+1..N]}$ for the right side,
		and the $\lambda_{\alpha_k}^{2}$ the eigenvalues of both $\rho^{[1..k]}$ and $\rho^{[k+1..N]}$, with $\lambda_{\alpha_k}\geq0$, $\sum_{\alpha_{k}=1}^{\chi_k}
		\lambda_{\alpha_k}^{2}=1$. The $\lambda_{\alpha_k}$ are the matrix elements $(\lambda^{[k]})_{\alpha_k}$ and either of the 
		$| \Phi_{\alpha_k} \rangle$ can be used to build the matrices $\Gamma^{[k]j_k}$ from equation (\ref{eq:mps_definition}).
		Thus, it is possible to keep only the largest $\chi_c$ eigenvalues and throw away
		the rest while re-normalizing the state, such that the sum of the discarded values is smaller than an arbitrary error $\epsilon$
		\begin{equation}
			\sum_{\alpha_{k}={\chi_c}}^{\chi_k} \lambda_{\alpha_k}^{2}<\epsilon .
		\end{equation}
		The best candidates are states for which the Schmidt dimension, and hence the dimension of
		the MPS matrices is low (like product states), or states for which the Schmidt coefficients display an exponential decay,
		where a large number of eigenvalues can be discarded without significant information loss.
		       
		The great advantage of the TEBD algorithm is the possibility to compute the time-evolution of a state using a time-dependent Hamiltonian
		\begin{equation}\label{eq:schroedinger_picture}
		| \Psi(t+ \delta t) \rangle = e^{{-\frac{i}{\hbar}H(t)\delta t}}| \Psi(t) \rangle .
		\end{equation}
		In a numerical implementation, ${\delta}t$ has to be discrete, such that the total simulation time $\tau={n_t}\cdot{\delta}t$, where $n_t$ 
		is the number of time steps and ${\delta}t$ the numerical time step. 
		The time-evolution can be used as well to calculate the ground state $\ket{\psi_{gr}}$ of a Hamiltonian $H$. This is done by taking the time to
		be imaginary and projecting (effectively `cooling') a starting state $\ket{\psi_P}$ to the ground state of $H$
		\begin{equation}
		    \ket{\psi_{gr}}=\lim_{\tau\rightarrow\infty}\frac{e^{-H\tau}\ket{\psi_P}}{||e^{-H\tau}\ket{\psi_P}||} .
		\end{equation}
		Given a one-dimensional system of size $N$ with nearest-neighbour interaction
		\begin{equation}
			H_N=\sum\limits_{l=1}^{N}K^{[l]}_1 + \sum\limits_{l=1}^{N}K^{[l,l+1]}_2
		\end{equation}
		we can split the Hamiltonian into two sums $H_N = F + G$ over even and odd sites, with 
		\begin{eqnarray}
			F \equiv \sum_{even \ \ l}(K^{l}_1 + K^{l,l+1}_2) = \sum_{even \ \ l}F^{[l]} ,\nonumber\\
			G \equiv \sum_{odd \ \ l}(K^{l}_1 + K^{l,l+1}_2) = \sum_{odd \ \ l}G^{[l]} .
		\end{eqnarray}	
		By using the Suzuki-Trotter decomposition for exponential operators
		\begin{eqnarray} 
			e^{{-\frac{i}{\hbar}H{\delta}t}} &=& e^{-\frac{i}{\hbar}(F+G){\delta}t}\nonumber\\
			 &=& e^{-\frac{i}{\hbar}\frac{F}{2}{\delta}t}e^{-\frac{i}{\hbar}G{\delta}t}e^{-\frac{i}{\hbar}\frac{F}{2}{\delta}t} + {{\cal{O}}}({{\delta}t}^3) ,
		\end{eqnarray}
		we can reduce the process of time-evolution to the successive application of operators which act only on two sites
		\begin{eqnarray}\label{eq:even_and_odd}
		    e^{-\frac{i}{\hbar}\frac{F}{2}{\delta}t} &=& \prod_{even \ \ l}e^{-\frac{i}{\hbar}{\frac{F}{2}}^{[l]}{\delta}t} , \nonumber\\
		    e^{-\frac{i}{\hbar}{G}{\delta}t} &=& \prod_{odd \ \ l}e^{-\frac{i}{\hbar}{{G}}^{[l]}{\delta}t} .
		\end{eqnarray}
		The TEBD algorithm then proceeds to update the MPS representation by calculating the new Schmidt decomposition at the corresponding bond each time after a 
		two-site operator is applied to it. This succession of two-site operators imposes a linear dependence of the computational cost on the system size $N$. 
		Each time a Schmidt decomposition is computed, we can truncate the dimension of the MPS by keeping only a maximal number of eigenvalues, using a suitable 
		limitation criterion. This is desirable, as the computational cost of both Schmidt decomposition and update of the MPS representation is 
		$\mathcal{O}({\chi_c}^3)$ as explicitly shown in
        Refs.~\onlinecite{Vidal03} and~\onlinecite{Vidal04}.
		The alternate updating of even and odd sites according to (\ref{eq:even_and_odd}) makes the TEBD algorithm highly parallelizable using up to
		$(N-1)$/2 threads with a very low communication overhead.
		
		\begin{figure}[t]
			\begin{center}
				\includegraphics[width=1\hsize]{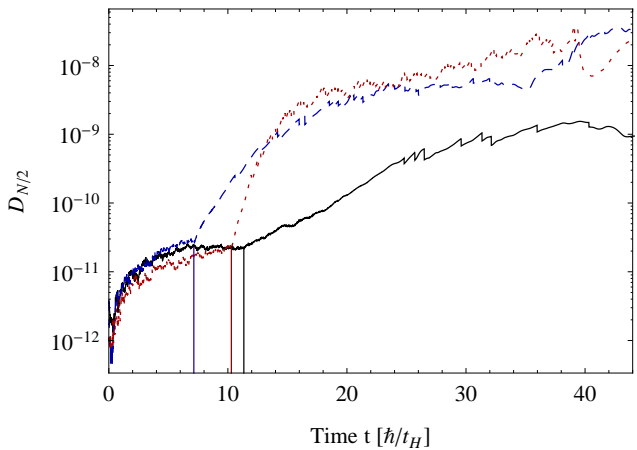}
			\caption{(Color online) Discarded weight $D_{N/2}$ for different parameters. solid black curve: $\Delta\epsilon = 0.5 t_{\text{H}}, V_{\text{H}} = 0.8 t_{\text{H}}$, 
			         dashed blue curve: $\Delta\epsilon = 1 t_{\text{H}}, V_{\text{H}} = 1.6 t_{\text{H}}$,
			         dotted red curve: $\Delta\epsilon = 2 t_{\text{H}}, V_{\text{H}} = 0$. 
			         The vertical lines indicate the runaway times described in the text.
				\label{fig:discarded_weight}}
			\end{center}
		\end{figure}

		One obvious error source is the one stemming from the discretization of time in order to numerically compute equation (\ref{eq:schroedinger_picture}),
		so it is necessary to keep the actual time step ${\delta}t$ small enough to have a good approximation of $H(t)$ in the interval $[t,t+{\delta}t]$.
		But this error source is not inherent to the TEBD implementation. The main error sources in the algorithm are the Suzuki-Trotter approximation and the 
		Schmidt truncation. In order to improve the time step error, one can use higher dimensional Suzuki-Trotter formulas, at a computational cost which scales 
		linearly with higher-order approximations\cite{Hatano05}, or we can decrease the time step ${\delta}t$, which also comes at a linear cost 
		$n_t=\frac{{\delta}t}{\tau}$. 
		The dominating error in probably all setups of interest is thus the truncation error. It is also very difficult to compensate for this error, because of the 
		$\mathcal{O}({\chi_c}^3)$ scaling of the computational cost. As a trivial example, when starting a real time evolution with a state with Schmidt dimension 
		smaller than $\chi_c$, the simulation runs with a constant truncation error.
		As the simulation continues, there comes a point where more states than $\chi_c$ are needed, and the truncation error quickly overcomes the 
		Trotter error, which basically defines a runaway time for the simulation, as reported in Ref.~\onlinecite{Gobert05}.
		A reasonable and often used estimator of the truncation error is the discarded weight
		\begin{equation}
			D_k = 1 - \sum_{\alpha_{k}=1}^{\chi_m} \lambda_{\alpha_k}^{2}
		\end{equation}
		where $D_k$ denotes the discarded weight for a bipartite split at the $k$-th bond and $\chi_m < \chi_c$ is the number of kept eigenvalues.
		For our real time simulations, we use a maximal Schmidt dimension in the range of $300 \le \chi_c \le 500$, depending on the specific parameter combination.
		We use a site-dependent $\chi_c(\text{site})$ and allow our simulations to adapt $\chi_c(\text{site})$ if test values (discarded weight, von-Neumann entropy)
		show the necessity to do so. We have to remark that for higher $|V_{\text{H}}|$ and $\Delta\epsilon$ the correlations (and thus the needed Schmidt dimension) 
		within the spinless fermion model grow very quickly.
		Hence, the chosen $\chi_c$-range can lead to significantly larger errors in the regime $|V_{\text{H}}| \ge 1.6 t_{\text{H}}$
		and $\Delta\epsilon \ge 3 t_{\text{H}}$.
		
		\begin{figure}[t]
			\begin{center}
				\includegraphics[width=1\hsize]{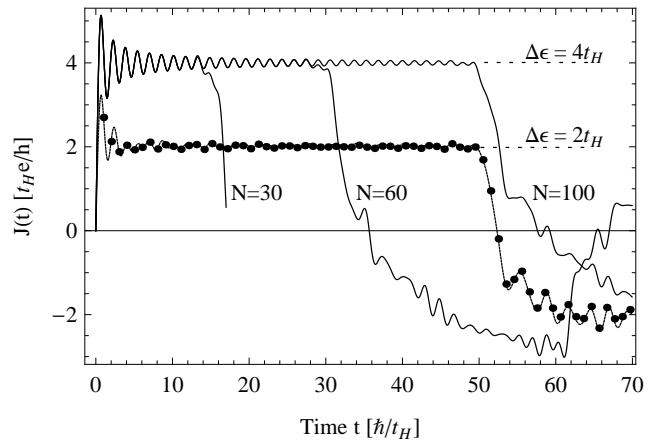}
				\caption{Current for non-interacting fermions for setup (I).
                Shown are TEBD (dots on top of the dashed line) and exact
                results (dashed and solid lines) obtained with 
								 the equation of motion method for the  single particle reduced density matrix (\ref{eq:one_body_definition}) for different system sizes $N$.
				\label{fig:exact}}
			\end{center}
		\end{figure}
		
	\begin{figure}[t]
		\begin{center}
			\includegraphics[width=1.0\hsize]{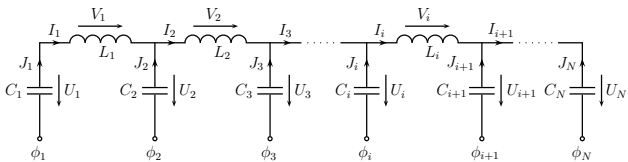}
			\caption{Classical $LC$ line with on-site applied potentials $\phi_i$.
			\label{fig:classical_setup}}
		\end{center}
	\end{figure}

		Figure~\ref{fig:discarded_weight} shows the discarded weight for different parameters for a split in the middle of the chain.
		One can see the runaway time indicated by the vertical lines.
		However, the maximal discarded weight over all simulations, sites and times up to $40\hbar/t_{\text{H}}$ we measured is yet smaller than $10^{-6}$.
		Our simulations reveal that when taking a maximal Schmidt dimension of $\chi_c = 600$ instead of 300, the runaway time is shifted only about 20\% to
		larger times and our quantities of interest are only slightly changed.
		Moreover, the stationary current computed with our method, which is described later, is only changed by less than 0.5\%.

		Figure~\ref{fig:exact} shows the very good agreement between our TEBD simulations and the exact calculation using the single particle 
		reduced density matrix (\ref{eq:one_body_definition}) for spinless fermions in the tight-binding model.
		We use a time step $\delta t=0.01$ and a typical simulation runs approximately 1/2 to 8 hours for a ground state calculation and about 4 hours to (seldomly) 
		1 week for a real time evolution. We use a multi-threaded version of TEBD with an extremely low overhead (less than 1\% for 10 processors
		 and less than 5\% for up to 96 processors).
		 The overall cost of a TEBD simulation is  slightly higher than that of
a DMRG calculation on a single processor. If several processors
are used in parallel the overhead of DMRG calculations become rapidly
prohibitive, exceeding 100 \% for as few as 4 processors.\cite{Hager04}
Thus TEBD is already twice as fast as DMRG for 4 processors and the difference is likely 
to increase rapidly with the number of processors.  We have used 12 threads for a TEBD 
simulation on average.
		
\section{Classical $LC$ line}

	In the following, our attention will be put on a classical model: the so-called $LC$ line, shown in Fig.~\ref{fig:classical_setup}.
	The $LC$ line has been under research as an electrical circuit implementation of a Toda chain\cite{Hirota73,Toda75,Singer99},
	a nonlinear oscillator chain that, among others, describes the propagation of soliton waves.
	Further investigations of the $LC$ line have so far been focused on the continuous and linear case.\cite{Ingold92}
	While this setup has therefore been analyzed for the nonlinear and the linear and continuous case, our focus lies
	on a discrete and linear model for finite and infinite sizes.
	
	In the following, we show that the $LC$ line serves well as the classical representation of a one-dimensional quantum wire described by 
	the tight-binding Hamiltonian, while our main focus lies on the behaviour of the current, especially 
	the influence of finite-size effects and stationary values for the infinite-size limit.
	We present solutions for a setup in which an initial imbalance of charge carriers
	on the condensators leads to an oscillating, rectangular current curve as the charge carriers move back and forth in the $LC$ line:
	the formerly described setup (I).
	Scaling the system size $N$ to infinity leads to an enlargement of the square wave period, and for large times a stationary current $\bar{I}$ is flowing.
	Subsequently, a method is presented to compute the stationary current $\bar{I}$ for infinitely large systems
	using finite-size simulation results. Based on the comparison of the classical and quantum mechanical model
	this approach will be applied to quantum systems.
	
	Figure~\ref{fig:classical_setup} shows the classical, electronic setup of a one-dimensional wire that possesses
	many properties of a one-dimensional quantum chain. This $LC$ line is a combination of condensators and inductors whereas the present
	model is extended to enable on-site applied potentials $\phi_i$.
	For this setup, the following relations can be derived using elementary electrotechnical relations and Kirchhoff's laws
	\begin{eqnarray}\label{eq:classical_laws}
	    J_i & = & - C_i \dot{U}_i  \quad  ;\quad i = 1,2,..,N\nonumber\\
	    V_i & = & L_i \dot{I}_i         ;\quad i = 1,2,..,N-1\\
	    I_{i+1} & = & I_i + J_{i+1}     ;\quad i = 0,1,..,N-1\nonumber
	\end{eqnarray}
	and
	\begin{equation}\label{eq:V_i}
	    V_i = U_i - U_{i+1} + \phi_{i} - \phi_{i+1} \quad ; \quad i = 1,2,..,N-1
	\end{equation}
	where $I_0=I_N=0$ was defined. $V_i$ in equation (\ref{eq:V_i}) results from the potential difference between the
	left ($\phi_i + U_i$) and right ($\phi_{i+1} + U_{i+1}$) connection of the $i$-th inductor.
	$C_i$ denotes the capacity of the $i$-th capacitor, $U_i$ the voltage drop over the capacitor, $L_i$ the inductance of the $i$-th
	inductor and $I_i$ and $J_i$ are the currents as shown in Fig.~\ref{fig:classical_setup}.
	$Q_i$ is the charge of the $i$-th capacitor and the $\phi_i$ denote externally applied potentials.
	From the equations above follows
	\begin{equation}\label{eq:matrix_dgl}
	    \ddot{\vec{I}} = - M \vec{I} + \dot{\vec{\Phi}}
	\end{equation}
	with
	\begin{eqnarray}
	    M_{ij} = &-& \delta_{i,j+1} \left( \frac{1}{L_i C_i} \right) - \delta_{i,j-1} \left( \frac{1}{L_i C_{i+1}} \right)\nonumber\\
	             &+& \delta_{ij} \left( \frac{1}{L_i C_{i+1}} + \frac{1}{L_i C_i} \right)
	\end{eqnarray}
	and $\dot{\Phi}_i = (\dot{\phi}_{i} - \dot{\phi}_{i+1})/L_i$.
	We consider an $LC$ line with equal capacities and inductances
	\begin{equation}
			L_i=L \quad \text{and} \quad C_i = C \quad \forall i
	\end{equation}
	and temporally constant applied potentials $\dot{\vec{\Phi}}=\vec{0}$ for which it follows
	\begin{equation}\label{eq:classical_ith_current}
			I_i(t) = \sqrt{\frac{2}{N}} \sum_{k=1}^{N-1} \sin\left(\frac{k i \pi}{N}\right)
							 \left[ b_k \sin\left(\omega_k t\right) + d_k \cos\left(\omega_k t\right)\right]
	\end{equation}				
	and
	\begin{equation}\label{eq:omega_k}
	    \omega_k = \frac{2}{\sqrt{LC}}\sin\left(\frac{k\pi}{2N} \right) .
	\end{equation}
	With $Q_{0,i} := Q_i(t=0)$ and $\phi_{0,i} := \phi_i(t=0)$ one has
	\begin{equation}\label{eq:dotI_0}
			\dot{I}_i (0) = \frac{1}{LC} \left( Q_{0,i} - Q_{0,i+1} \right) + \frac{1}{L} \left( \phi_{0,i} - \phi_{0,i+1} \right)
	\end{equation}
	and thus
	\begin{eqnarray}\label{eq:b_k2}
			b_k &=& \frac{\sqrt{2}}{\omega_k L \sqrt{N}} \sum_{i=1}^{N-1} \left[ \frac{1}{C}(Q_{0,i} - Q_{0,i+1}) \right. \nonumber\\
			    & &\quad\quad \left.\phantom{\frac{|}{|}} + \phi_{0,i} - \phi_{0,i+1} \right] \sin\left(\frac{k i\pi}{N}\right) .
	\end{eqnarray}

	\subsection{Stationary current}
		
		In the following, we assume that initially (at $t=0$) no current is flowing
		\begin{equation}
				I_i(0)=0\quad\forall i \quad \Rightarrow \quad d_k = 0\quad\forall k
		\end{equation}
		and that the chain is divided into two halves in which the charge carriers are uniformly distributed, i.e.
		\begin{eqnarray}\label{eq:theta_like_charge_distribution}
				Q_L = Q_{0,i=1,..,\frac{N}{2}} \quad, \quad & & Q_R = Q_{0,i=\frac{N}{2}+1,..,N} \nonumber\\
				\text{and} \quad & & \phi_{i}(t) = 0 \quad \forall i,t ,
		\end{eqnarray}
		which corresponds to the previously mentioned setup (I).
		
		\begin{figure}[t]
			\begin{center}
				\includegraphics[width=1\hsize]{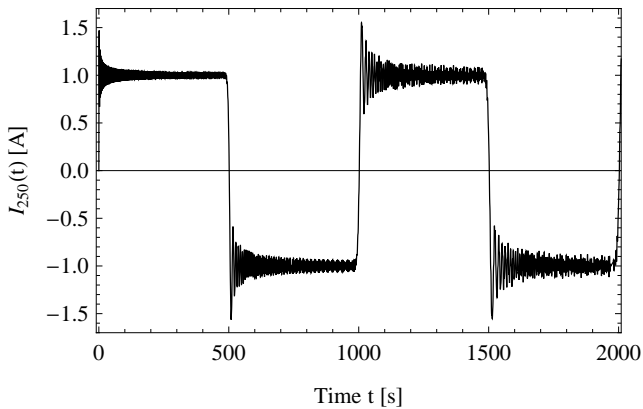}
				\caption{Current $I_{250}(t)$ for a classical wire of size $N=500$ through the
		 		    		 inductor at position $N/2$. $L=C=1$, $Q_L=Q_{0,i \le N/2}=1$ and $Q_R=Q_{0,i > N/2}=-1$
				\label{fig:classical_current_example}}
			\end{center}
		\end{figure}		
		
		Figure~\ref{fig:classical_current_example} shows the current through the inductor $L_{N/2}$ according to
		the general formula for the current at site $N/2$ for any even $N$
		\begin{equation}\label{eq:current_finite}
				I_{N/2}(t) = \frac{Q_L - Q_R}{N\sqrt{LC}} \sum_{k=1}^{N/2}\frac{\sin\left(\frac{2}{\sqrt{LC}}\eta_k t\right)}{\eta_k}
		\end{equation}
		with
		\begin{equation}\label{eq:eta_and_omega}
				\eta_k = \omega_{2k-1} \frac{\sqrt{LC}}{2} = \sin\left( \frac{(2k-1)\pi}{2N} \right) .
		\end{equation}
		All values for the classical system ($Q_i$, $t$, $L$, $C$, etc.) are given in S.I. units unless otherwise specified.
		
		While the rectangular shape of the dominant oscillation is understandable through the picture of charge carriers moving back and forth in the line,
		the rapid oscillation on top of the square wave is harder to explain and we will later take a closer look at it.
		The period $T^{\text{max}}_{\text{cl}}$ of the square wave can be calculated
		by using equation (\ref{eq:current_finite}) and (\ref{eq:eta_and_omega}) and taking only the smallest frequency into account
		\begin{equation}\label{eq:T_cl}
				T^{\text{max}}_{\text{cl}} = \frac{\pi\sqrt{LC}}{\sin\left(\frac{\pi}{2N}\right)} .
		\end{equation}
		For large system sizes ($N \gg 1$) this period is approximately given by
		\begin{equation}\label{eq:approx_T_max}				
		  	 T^{\text{max}}_{\text{cl}} \approx 2N\sqrt{LC}
		\end{equation}
		and is thus linearly dependent on $N$.
		In the second half period of the current in Fig.~\ref{fig:classical_current_example}
		one can see a beat upon the rapid oscillation. This is due to the unequally distributed charges which
		form an oscillating pattern and so create additional frequencies in the current.
		This effect also occurs in quantum wires, not only after the reflection at the right border but already
		from the beginning due to using an initial (ground) state with
        non-uniform local densities.
		
		In contrast to the oscillating current in a finite system where the charge carriers are scattered at the borders,
		we assume that the current in an infinitely large system is constantly flowing in one direction.
		To validate this assumption one has to execute the limit $N\rightarrow\infty$ first, and then take the limit $t\rightarrow\infty$.
		In the limit $N\rightarrow\infty$ one gets from equation (\ref{eq:current_finite})
		\begin{eqnarray}\label{app:ext_sine_integral_firstpart}
				I_{\infty}(t) &=& \lim_{N\rightarrow\infty} I_{N/2}(t) \nonumber\\
				              &=& \frac{Q_L - Q_R}{\sqrt{LC}} \frac{1}{\pi} \int_{0}^{1}
						    					\frac{\sin\left(a \xi \right)}{\xi\sqrt{1-\xi^2}} d\xi
		\end{eqnarray}
		with $a = 2t / \sqrt{LC}$ and $\varphi = 2 / \sqrt{LC}$.
		The solution is given by
		\begin{eqnarray}\label{eq:i(t)}
				I_{\infty}(t) &=& \frac{\left(Q_L-Q_R\right)t}{2LC} \bigl[\BesselJ_0\left(\varphi t\right) \left(2-\pi\StruveH_1\left(\varphi t\right)\right)\nonumber\\
						    & &\quad\quad +~ \pi \BesselJ_1\left(\varphi t\right) \StruveH_0\left(\varphi t\right)\bigr]
		\end{eqnarray}
		where $\BesselJ_n(x)$ are the Bessel functions of the first kind and $\StruveH_n(x)$ denote the Struve functions.\cite{Abramowitz72}
		
		\begin{figure}[t]
			\begin{center}
				\includegraphics[width=1\hsize]{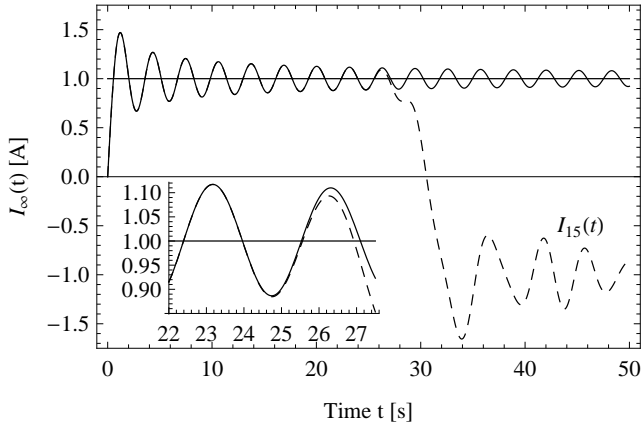}
				\caption{Solid curve: Current $I_{\infty}(t)$ for a classical infinite wire through the
		    				 inductor at the middle with $L=C=1$, $Q_L=1$ and $Q_R=-1$. The dashed curve is the current for a finite system with size $N=30$.
		    				 The constant line denotes the stationary value $1$ against which the solid line converges.
				\label{fig:infinite_classical_current}}
			\end{center}
		\end{figure}		
		
		The current for the infinite system (\ref{eq:i(t)}) is shown in Fig.~\ref{fig:infinite_classical_current} and it shows that
		for smaller times the curve for finite system sizes matches the one for the infinitely large LC line.
		Hence, a very accurate approximation of the value of a stationary current in an infinitely large system can be extracted
		from the value of a corresponding finite system result. This behaviour will be later used for the analysis of quantum systems.
		
		Using the approximations (\ref{eq:approx_bessel_struve}) for the Bessel and Struve functions
		we find the value of the stationary current in the infinitely large system
		\begin{equation}\label{eq:infinite_current}
				\bar{I}_{\text{cl}} = \lim_{t\rightarrow\infty} I_{\infty}(t) = \frac{Q_L - Q_R}{2 \sqrt{LC}} .
		\end{equation}
		The conductance $G_{\text{cl}}$ for the stationary case is given by
		\begin{equation}\label{eq:g_cl}
				G_{\text{cl}} = \frac{\bar{I}_{\text{cl}}}{V} = \frac{1}{2}\sqrt{\frac{C}{L}}
		\end{equation}
		with a local voltage drop $V = \left(Q_L - Q_R\right)/C$ in the middle of the chain.

	\subsection{Finite-time and finite-size effects}
	
		In order to describe finite-time effects in the infinite system we seek for a simpler current expression which does not only
		give the correct stationary value but also a good approximation for the short-time behaviour.
		Using the asymptotic series expansions (\ref{eq:asymptotic_approximations}), $I_{\infty}(t)$ can be transformed into the following form for $t\gg 1$
		\begin{eqnarray}\label{eq:I_gg_final}
				I_{\text{app}}(t) &= \bar{I}_{\text{cl}} &+ I_{\text{dev}}(t)\nonumber\\
													&= \bar{I}_{\text{cl}} &+ \frac{Q_L - Q_R}{16} \sqrt{\frac{\varphi}{\pi t}} \bigl[ 
											   (10\pi-32)(\sin^2(\varphi t)\nonumber\\
									  & &+~ \sin(\varphi t)\cos(\varphi t))\nonumber\\
										&	&+~ 4(\sin(\varphi t)-\cos(\varphi t))\bigr] + \mathcal{O}\left(\frac{1}{t}\right)
		\end{eqnarray}
		with $\varphi= 2/\sqrt{LC}$.
		$I_{\text{dev}}(t)$ denotes the deviation from the stationary current $\bar{I}_{\text{cl}}$ from equation (\ref{eq:infinite_current}).
		The order $\mathcal{O}\left(t^{-1}\right)$ of the rest term is only valid if the approximation of $\StruveH_1(x)$ in (\ref{eq:asymptotic_approximations})
		has an error of smaller order which presumably holds according to Ref.~\onlinecite{Aarts03}.
		
		\begin{figure}[t]
			\begin{center}
				\includegraphics[width=1\hsize]{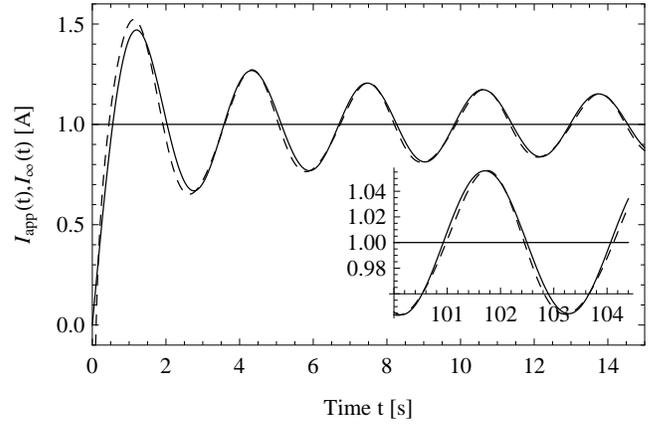}
				\caption{Exact current $I_{\infty}(t)$ (solid line) and
                approximative current $I_{\text{app}}(t)$ (dashed line)
			    			 with $L=C=1$, $Q_L=1$ and $Q_R=-1$.
				\label{fig:app_exact_current}}
			\end{center}
		\end{figure}		
		
		As shown in Fig.~\ref{fig:app_exact_current}, expression (\ref{eq:I_gg_final}) is a pursuasive approximation even for smaller times
		and thus a good explanatory basis for short-time effects in the $LC$ line.		
		It states a general $\mathcal{O}(1/\sqrt{t})$ decay of the amplitudes of the rapid oscillation in the current curve
		which is the maximal order of the error for an approximation of the stationary current (\ref{eq:infinite_current}) using a finite-time (i.e.
		a finite-size) result.

		Instead of looking at the current through a single inductor, one can consider the	current through two or more neighbouring ones.
		We will show that the short time behaviour of this current has some advantages compared to the simple $I_{N/2}(t)$ curve.
		We define the quantity of interest for even $N$ as
		\begin{equation}
			I_{\text{m,}N/2}(t) = \frac{1}{2} I_{N/2}(t) + \frac{1}{4} \left( I_{N/2+1}(t) + I_{N/2-1}(t) \right)
		\end{equation}
		which is equal to the expression $\frac{1}{2} ( I_{N/2}(t) + I_{N/2+1}(t) )$,
		whereas for the latter expression the subsequently explained effect also occurs for odd $N$ for the substitution $N/2 \rightarrow (N-1)/2$.
		In the limit $N \rightarrow \infty, t \rightarrow \infty$, $I_{\text{m,}N/2}(t)$ converges against the same value as $I_{N/2}(t)$
		which implies that the stationary current is as well assessable via the current through two or more adjacent inductors.
		From (\ref{eq:classical_ith_current}) one gets
		\begin{eqnarray}
			I_{\text{m,}N/2}(t) &=& \frac{Q_L - Q_R}{NLC} \sum_{k=1}^{N-1} \frac{1}{\omega_k} \sin\left(\omega_k t\right)
																			\sin^2\left(\frac{k \pi}{2}\right) \nonumber\\
																	& & \cdot \left[1 + \cos\left(\frac{k \pi}{N}\right) \right]
		\end{eqnarray}
		with expression (\ref{eq:b_k2}) for $b_k$ and the step distribution for the charges (\ref{eq:theta_like_charge_distribution}).
		In the limit $N \rightarrow \infty$ it follows
		\begin{equation}\label{eq:dot_current_integral}
			I_{\text{m,}\infty}(t) = \frac{Q_L - Q_R}{\pi\sqrt{LC}} \int_{0}^{\frac{\pi}{2}} 
														   \frac{\sin\left(\frac{2}{\sqrt{LC}}\sin\left( x \right) t\right)}{\sin\left( x \right)} \cos^2(x)dx .
		\end{equation}
		The integral can be solved in analogy to (\ref{app:ext_sine_integral_firstpart}) with the substitution $\xi = \sin\left( x \right)$ and gives
		\begin{eqnarray}\label{eq:I_dot_infinity}
				I_{\text{m,}\infty}(t)
						 &=& \frac{Q_L - Q_R}{\sqrt{LC}} \left[ \BesselJ_1(\varphi t) \left(\frac{\pi\varphi t}{4}\StruveH_0(\varphi t)
						 		 -\frac{1}{\varphi^2 t^2} - \frac{1}{2}\right) \right.\nonumber\\
						 & & +~ \left.\BesselJ_0(\varphi t) \left(\frac{\varphi t}{2} + \frac{1}{2\varphi t} - \frac{\pi \varphi t}{4} \StruveH_1(\varphi t)\right)\right] .
		\end{eqnarray}
		
		\begin{figure}[t]
			\begin{center}
				\includegraphics[width=1\hsize]{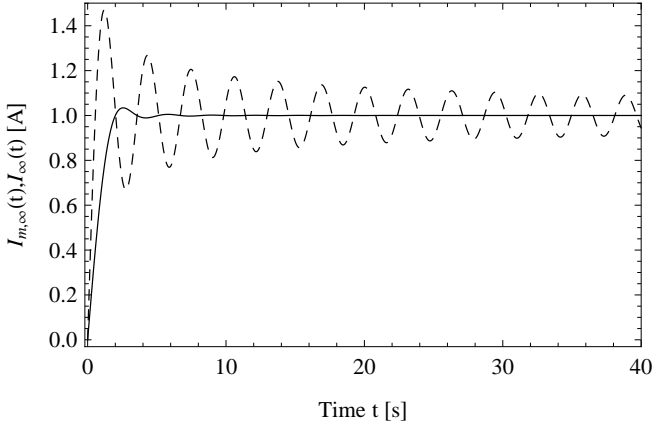}
				\caption{Solid curve: Current $I_{\text{m,}\infty}(t)$ through two adjacent inductors in the middle;
								 equation (\ref{eq:I_dot_infinity}). Dashed curve: Current $I_{\infty}(t)$ from
		    				 equation (\ref{eq:i(t)}) through a single inductor. $L=C=1$, $Q_L=1$ and $Q_R=-1$.
				\label{fig:compare_single_inductor_with_double}}
			\end{center}
		\end{figure}		
		
		This result, plotted in Fig.~\ref{fig:compare_single_inductor_with_double} together with equation (\ref{eq:i(t)}) for the current $I_{\infty}(t)$ 
		through a single inductor, shows almost no rapid oscillations.		
		An approximative expression for the rapid oscillations $I_{\text{r}}(t) \approx I_{\text{dev}}(t)$
		can therefore be derived using the difference between the two current expressions (\ref{eq:I_dot_infinity}) and (\ref{eq:i(t)})
		\begin{eqnarray}\label{eq:I_osc_exact}
				I_{\text{r}}(t) 
							&=& I_{\text{m,}\infty}(t) - I_{\infty}(t)\nonumber\\
							&=& \frac{Q_L - Q_R}{4} \left[ \frac{\BesselJ_0(\varphi t)}{t} - \left(\frac{2}{\varphi t^2} + \varphi\right) \BesselJ_1(\varphi t) \right]
		\end{eqnarray}
		and using the approximations (\ref{eq:asymptotic_approximations}) for the remaining Bessel functions one has
		\begin{eqnarray}\label{eq:I_small_osc}
				I_{\text{r,app}}(t) &=& \frac{(Q_L - Q_R)}{32 \sqrt{\pi} \left(\varphi t\right)^{5/2} t}
						                    \left [ \left (5 (\varphi t)^2 + 8 (\varphi t)^3 \right.\right.\nonumber\\
						                & & \left.\left. +~ 15 \varphi t - 6 \right ) \cos(\varphi t) \left (5 (\varphi t)^2 - 8 (\varphi t)^3 \right.\right.\nonumber\\
						                & & \left.\left. -~ 15 \varphi t -6
                                        \right ) \sin(\varphi t)  \right] .
		\end{eqnarray}
		Hence the rapid oscillations have a radial frequency $\varphi=2 / \sqrt{LC}=2\pi / T^{\text{min}}_{\text{cl}}$ and thus
		\begin{equation}\label{eq:T_min_cl}
				T^{\text{min}}_{\text{cl}} = \pi\sqrt{LC} .
		\end{equation}
		
	\subsection{\label{subsection:method}Determining the stationary current from finite system values}
		
		There are various methods to calculate the stationary current from simulation results stemming from systems with finite size.
		The most obvious and easiest way to do so is to use the curve of a current through the middle inductor
		(or the mean value of currents through two or more adjacent inductors) and to compute the mean value over a certain time interval
		in the quasi-stationary regime.
		In contrast to this approach, we will present a more analytical approach using the Fourier transformation.
		Although one cannot prove that the following strategy can be applied to quantum mechanical systems
		with complicated interactions, several numerical and theoretical indications are given in sections \ref{section:finite-size-effects-and-stat-current}
		and \ref{section:spinless-fermion-model}.
		We start with a square wave described by its Fourier series
		\begin{equation}\label{eq:rectangular_curve}
				\mathcal{R}_m(t) := \frac{4A}{\pi} \sum_{k=1}^{m} \frac{\sin\left((2k-1) \chi t\right)}{2k-1}
		\end{equation}
		where $m$ determines the number of harmonics, $A$ denotes the value of the quasi-stationary plateau
		and $\chi=2\pi / T^{\text{max}}_R$ where $T^{\text{max}}_R$ is the period of the square wave. 
		This signal has certain similarities with the current curve $I_{N/2}(t)$.
		It possesses a quasi-stationary regime which becomes infinite for $\chi \rightarrow 0$ (equals $N \rightarrow \infty$ for the $LC$ line)
		and shows a rapid and decaying oscillation on top of it.
		The main difference between the composition of the current curve (\ref{eq:current_finite}) and the square wave 
		is the presence of sine-functions that enclose the $(2k-1)$-terms.
		
		The current in (\ref{eq:current_finite}) is dependent on $\eta_k$ and it is thus convenient to write $I_{N/2}^{[\eta_k]}(t)$
		where $\eta_k = \sin(\frac{(2k-1)\pi}{2N})$.		
		Choosing an $\eta_k$ without the enclosing sine-functions gives
		\begin{equation}\label{eq:I_functional}
				I_{N/2}^{\left[\frac{(2k-1)\pi}{2N}\right]}(t) = \frac{2(Q_L - Q_R)}{\pi\sqrt{LC}} \sum_{k=1}^{N/2}\frac{\sin\left((2k-1) \varphi' t\right)}{2k-1}
		\end{equation}
		with $\varphi'=\varphi\pi / (2N) = \pi / (N\sqrt{LC})$.
		Since expression (\ref{eq:I_functional}) now describes a square wave, one can compare it with (\ref{eq:rectangular_curve}) and get
		its quasi-stationary value
		\begin{equation}\label{eq:A_equals_Icl}
				A = \frac{Q_L - Q_R}{2 \sqrt{LC}} = \bar{I}_{\text{cl}}
		\end{equation}
		which is equal to the value that was derived for the stationary current in equation (\ref{eq:infinite_current}).
		Remarkably the enclosing sine-functions do not change the stationary value, i.e. the expression (\ref{eq:I_functional})
		and the current in an $LC$ line have the same stationary value for $N\rightarrow\infty$.
		In case of expression (\ref{eq:I_functional}) this value is equal to the quasi-stationary value (\ref{eq:A_equals_Icl})
		which leads to the idea to calculate the quasi-stationary value from a given current signal in a finite system (using the procedure described below)
		and use it as approximation for the stationary current value for $N\rightarrow\infty$.

		Applying the Fourier-transformation
		\begin{equation}\label{eq:Fourier_transform}
				\tilde{I}_{N/2}(\omega) := \int_{-\infty}^{\infty} I_{N/2}(t) e^{i \omega t} dt
		\end{equation}
		term by term to expression (\ref{eq:current_finite}) gives
		\begin{equation}\label{eq:ft_classical}
				\tilde{I}_{N/2}(\omega) = i \cdot \frac{(Q_L - Q_R)}{N\sqrt{LC}} \sum_{k=1}^{N/2} \frac{
						\delta(\omega+\varphi\eta_k) - \delta(\omega-\varphi\eta_k)
				}{\eta_k}
		\end{equation}	
		with equation (\ref{eq:eta_and_omega}) for $\eta_k$	and $\varphi=2 / \sqrt{LC}$.
		If numerical data with $m$ values are given, one can only apply a discrete Fourier transformation which preserves the
		area underneath a delta peak, such that for any interval $\Delta t$ containing a single delta function $a\cdot\delta(x-x_0)$
		the peak is of height $a / \Delta t$.
		Taking only the first ($k=1$) and only the positive delta-peak into account and using the discrete transformation
		\begin{equation}
				\tilde{f}(j) = \frac{1}{m} \sum_{k=0}^{m-1} f(k) e^{-i k j \omega_0} \quad \phantom{=}\text{with} \phantom{=}\omega_0 = \frac{2\pi}{m}
		\end{equation}		    
		and $j=0,1,...,m-1$, one gets for the current
		\begin{equation}
				\bar{I}_{\text{cl}}=\frac{\text{Im}[\tilde{I}_{N/2}(\omega_1)] N \Delta t}{2} \cdot \sin\left(\frac{\pi}{2N}\right)
		\end{equation}
		where $\text{Im}[\tilde{I}_{N/2}(\omega_1)]$ is the height of the first discrete peak ($k=1$).
		In fact, one cannot expect to know the precise form of $\eta_k$ for complicated quantum systems. Instead, one can take
		a first order approximative expression.
		For $N \gg 1$ it holds $\sin\left(\frac{\pi}{2N}\right) \approx \frac{\pi}{2N}$, which implies that the given curve is a square wave, and thus
		\begin{equation}\label{eq:approximative_stationary_current}
				\bar{I}_{\text{cl}} \approx \frac{\pi}{4} \text{Im}[\tilde{I}_{N/2}(\omega_1)] \Delta t
		\end{equation}
		with a relative error\cite{Abramowitz72}
		\begin{equation}\label{eq:error_for_LC-Line_fft}
				\frac{|\Delta\bar{I}_{\text{cl}}|}{|\bar{I}_{\text{cl}}|} \le \frac{\pi^2}{8 N^2} .
		\end{equation}
		It is important to remark that equation (\ref{eq:approximative_stationary_current}) is not only an exact result for a square wave,
		but also an excellent approximation for the $LC$ line current with maximal error according to equation (\ref{eq:error_for_LC-Line_fft}).
		
\begin{figure}[t]
			\begin{center}
				\includegraphics[width=1\hsize]{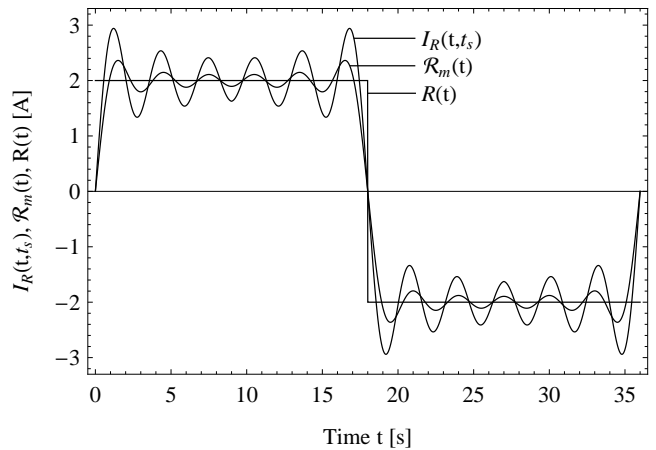}
				\caption{Constructed signal $I_{\mathcal{R}}(t,t_s)$ from simulation results together with the Fourier series representation 
								 of a square wave $\mathcal{R}_m(t)$ from (\ref{eq:rectangular_curve}) with $m=8$ and an ideal square wave $R(t)$. 
		    				 The Fourier transformation of  $I_{\mathcal{R}}(t,t_s)$ and expression (\ref{eq:approximative_stationary_current}) was used to 
		    				 calculate the amplitude of $R(t)$.
				\label{fig:rectangle_curves}}
			\end{center}
		\end{figure}

		As our simulations reveal, the charge carriers in the $LC$ line coming from the
		left half first reach the right border at $t \approx T^{\text{max}}_{\text{cl}}/4$.
		After this time the finite system size has a significant influence on the simulation results.
		Applying a discrete Fourier transformation means to assume that a periodic function is given. 
		For that, and to minimize the upcoming error for times larger than $T^{\text{max}}_{\text{cl}}/4$ the following procedure is proposed:
		From a given current curve $I_{N/2}(t)$ stemming from simulation results, first determine a position $t=t_s$ in the vicinity of
		$T^{\text{max}}_{\text{cl}}/4$ --
		but smaller than $T^{\text{max}}_{\text{cl}}/4$
		-- where the curve could be mirrored along a parallel to the $y$-axis such that the loss of continuity is minimized, which
		is the case for instance at inflection points.
		Then consider a $4t_s$-periodic function, constructed as follows
		\begin{equation}\label{eq:4ts_function}
				I_{\mathcal{R}}(t, t_s) := 
					\left\lbrace
					\begin{array}{c c}
							I_{N/2}(t)\phantom{2t_s.....} 					&\text{for } 0   \le t  <   t_s\phantom{+..}\\
					 		I_{N/2}(2t_s - t) 	&\text{for } t_s \le t  <  2t_s\phantom{.,}\\
					 		- I_{N/2}(t - 2t_s)   &\text{for }2t_s \le t  <  3t_s\\
					 		- I_{N/2}(4t_s - t)   &\text{for }3t_s \le t \le 4t_s
					\end{array}
					\right.  .
		\end{equation}
		We then apply the Fourier transformation to the constructed signal (\ref{eq:4ts_function}) and use equation (\ref{eq:approximative_stationary_current})
		to calculate the corresponding stationary value of the current, see Fig.~~\ref{fig:rectangle_curves}.

	\subsection{A final note on setups}
		
		Instead of initially dividing the system into two halves with different charges $Q_L$ and $Q_R$, we can
		distribute the charge carriers homogeneously over the chain, apply a temporally constant voltage ($\phi_{0,L} - \phi_{0,R}$)
		and get the same result for the current as a function of time, which can be seen in equation (\ref{eq:b_k2}) for $b_k$.
		The reason for this lies in the linear character of our model which is as a consequence independent from the chosen setup.
		This is the main reason why we cannot fully compare the $LC$ line with a quantum system, since preparing the classical system in
		setup (II) gives the same outcomes as for setup (I), in contrast to our quantum simulation results.
		
\section{\label{section:finite-size-effects-and-stat-current}Finite-size effects and stationary current in quantum systems}

	\subsection{Finite-size induced oscillation}

		\begin{figure}[t]
			\begin{center}
				\includegraphics[width=1\hsize]{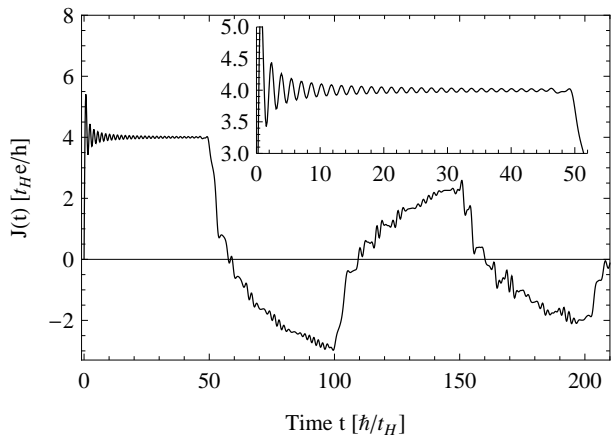}
				\caption{Current in the tight-binding model for a long time scale, calculated using the
		    				 one-particle equation of motion. The initial state was prepared with
								 a completely filled left half, a completely empty right half and $t_{\text{H}}=0$ for all bonds.
				\label{fig:tb_long_time}}
			\end{center}
		\end{figure}
		
		With both setups (I) and (II) one observes qualitatively similar currents in the tight-binding model
		(the spinless fermion model without interaction $V_{\text{H}} = 0$) as a function of time, as 
		shown in Figs.~\ref{fig:exact} and \ref{fig:tb_long_time}.
		First, there is a small transient regime for $t < t_a \lesssim 3h / t_{\text{H}}$
		with very rapid and dominant small oscillations.
		For long times $t > t_b \gg T^{\text{max}}$ the current becomes very irregular because of the progressive
		dephasing of moving particles. 
		Between $t_a$ and $t_b$ we observe an approximately rectangular wave with a period $T^{\text{max}}$ which 
		diverges with increasing system length (the corresponding leading frequency $\omega^{\text{max}} = 2\pi/T^{\text{max}}$ converges
		towards 0 in the thermodynamic limit), see Fig.~\ref{fig:tb_long_time}.
		The period of the rectangular oscillation is given by (Appendix~\ref{app:period})
		\begin{eqnarray}\label{eq:T_qm_in_text}
			T^{\text{max}} \approx \hbar \cdot \frac{N}{t_{\text{H}}} \quad \text{for} \quad N \gg 1 .
		\end{eqnarray}
		
		For spinless fermions, using a semi-classical picture, particles first flow from one side of the system to the opposite side because of the inhomogeneous
		density (first setup) or the potential difference (second setup). 
		Then they are reflected by the hard wall represented by the chain edge. 
		As there is no dissipation in our model, all reflected particles flow back with the same velocity 
		in the opposite direction until they reach the other chain edge and are again reflected and so on. 
		
		The progressive degradation of the rectangular signal can be understood using the same picture. 
		First, all particles flow in the same direction but, as they have different velocities,  
		they progressively come out of phase.
		For long times $t \gg t_b$, which can be checked up to the numerical double limit
		$t\approx 10^{300}$ for non-interacting fermions ($V_{\text{H}}=0$), our simulations show
		that the current does not go to zero but continues to oscillate with a period $T^{\text{max}}$.
		This can be understood in the picture of the classical $LC$ line as well as 
		for spinless fermions through the dominant amplitude of the $\omega^{\text{max}}$ oscillation.
		It is basically the same effect as if all contributions to a series representation of a square wave (\ref{eq:rectangular_curve})
		would `randomly' come out of phase. In that case, the contribution with the lowest frequency $\omega^{\text{max}}$, 
		which is the frequency of the rectangular oscillation, determines the frequency of the remaining irregularly shaped curve.

	\subsection{Rapid oscillation on top of the square wave}

		\begin{figure}[b]
			\begin{center}
				\includegraphics[width=1\hsize]{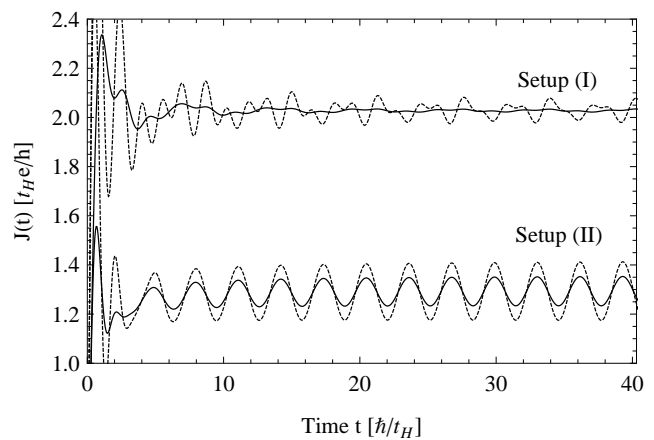}
				\caption{Current in the tight-binding model for $\Delta\epsilon=2t_{\text{H}}$,
		    				 calculated using the one-particle equation of motion.
		    				 The dashed lines are currents through a single bond in the chain while the solid ones represent the mean value of the currents
		    				 through two adjacent bonds in the middle.
				\label{fig:tb_dot_vs_bond}}
			\end{center}
		\end{figure}

		Setting the two expressions (\ref{eq:T_cl}) and (\ref{eq:T_qm_in_text}) for the periods of the classical and quantum mechanical system
		equal one gets
		\begin{equation}\label{eq:param_correspondance_1}
				t_{\text{H}} \approx \frac{\hbar}{2\sqrt{LC}} \quad \text{for } N \gg 1 .
		\end{equation}
		Further comparison with equation (\ref{eq:T_min_cl}) gives for the period of the rapid oscillation
		\begin{equation}\label{eq:T_min_qm}
				T^{\text{min}} = \pi\sqrt{LC} \approx \frac{h}{4t_{\text{H}}} \quad \text{for } N \gg 1 .
		\end{equation}
		
		Instead of preparing the system in a ground state of a Hamiltonian with applied potentials, 
		one can initially decouple all sites and fill the left half of the system with (uncorrelated) particles, 
		leaving the right half empty.
		This almost corresponds to choosing $\Delta\epsilon = 4 t_{\text{H}}$, insofar as a ground state with this applied voltage and coupled reservoirs 
		still has some correlations and unequally distributed on-site particle densities in it.
		The resulting current curve in Fig.~\ref{fig:tb_long_time} in the first half period closely resembles the curve shown in
		Fig.~\ref{fig:classical_current_example} for the $LC$ line.

		The magnification in Fig.~\ref{fig:tb_long_time} shows a rapid oscillation with a period according to equation (\ref{eq:T_min_qm}),
		in contrast to the oscillation shown in Fig.~\ref{fig:tb_dot_vs_bond} for setup (I) which is `disturbed' by oscillations 
		stemming from correlations and an unequal
		distribution of particles in the ground state due to a lower applied voltage $\Delta\epsilon < 4t_{\text{H}}$.
		Although the oscillation in Fig.~\ref{fig:tb_dot_vs_bond} for setup (II) is very regular, it does not stem from the same `classical' origin,
		which is confirmed by the fact that on the one hand two adjacent currents do not cancel out and on the other hand
		the oscillation does not fulfill equation (\ref{eq:T_min_qm}).
		
		Since in the following we are more interested in the stationary current flowing from an infinitely large source (the left half of the chain) 
		to a an infinitely large drain (the other half), we reference to 
		Ref.~\onlinecite{Wingreen93} for a discussion of time-dependent currents and
		to Ref.~\onlinecite{Branschaedel10} for a 
		detailed analysis  of short-time and finite-size effects (in the IRLM).

	\subsection{Applicability of the extrapolation approach to quantum systems}
		
		Although we have seen the close connection between the tight-binding system for initially uncorrelated particles and
		the classical $LC$ line, there exists a major difference in the behaviour of the respective currents: the damping of the
		rapid oscillations.
		While for the classical $LC$ line a $\mathcal{O}(t^{-\frac{1}{2}})$ dependence of the amplitudes according to equation (\ref{eq:I_gg_final})
		is predicted, this decrement is different for a quantum system.

		\begin{figure}[tb]
			\begin{center}
				\includegraphics[width=1\hsize]{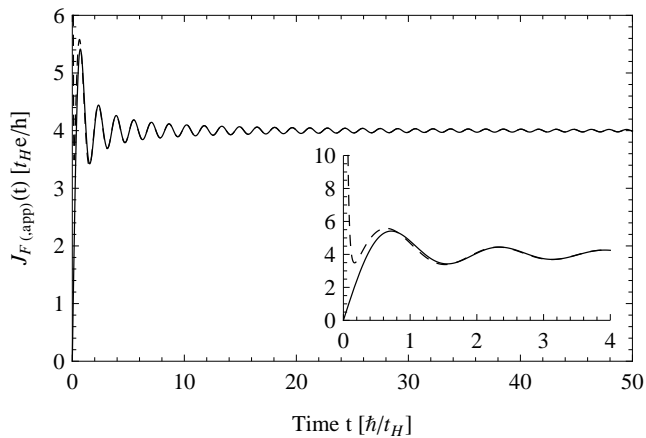}
				\caption{Solid line: $J_{\text{F}}(t)$ from equation (\ref{eq:dmu=4_current_from_hungarians}),
							   dashed line: corresponding approximation $J_{\text{F,app}}(t)$ from (\ref{eq:approx_dmu=4_current_from_hungarians}).
				\label{fig:exact_current_hungarians}}
			\end{center}
		\end{figure}

		The expectation value of the current for setup (I) and for initial conditions like in Fig.~\ref{fig:tb_long_time} (one completely filled and
		one completely empty half) is for $N\rightarrow\infty$ given by\cite{Antal99,Hunyadi04} (Appendix~\ref{app:current_in_quantum_system})
		\begin{equation}\label{eq:dmu=4_current_from_hungarians}
				J_{\text{F}}(t) = \frac{et_{\text{H}}}{\hbar} \omega t \left[ (\BesselJ_0(\omega t))^2 + (\BesselJ_1(\omega t))^2 \right] .
		\end{equation}
		with $\omega = 2 t_{\text{H}}/\hbar$.
		Using the asymptotic series expansions (\ref{eq:asymptotic_approximations}) yields for the current
		\begin{eqnarray}\label{eq:approx_dmu=4_current_from_hungarians}
				J_{\text{F,app}}(t) &=& \frac{et_{\text{H}}}{16h(\omega t)^2} \bigl[5-32 \omega t \cos(2\omega t) \nonumber\\
				                    & &+~ 4 \sin(2\omega t)\bigr] + \frac{4et_{\text{H}}}{h}
		\end{eqnarray}
		which is plotted together with expression (\ref{eq:dmu=4_current_from_hungarians}) in Fig.~\ref{fig:exact_current_hungarians}.
		
		The curve highly coincides with the one shown in Fig.~\ref{fig:tb_long_time} for $t < T^{\text{max}}/4$ which confirms
		the assumption that the borders in the finite quantum system for setup (1) only significantly 
		change the current after a time $T^{\text{max}}/4$, determined by the velocity of the particle density front wave\cite{Hunyadi04} 
		which moves from the middle of the chain to the borders.
		The quasi-stationary value $\bar{J} = 4et_{\text{H}}/h$ from (\ref{eq:approx_dmu=4_current_from_hungarians}) suits equation (\ref{eq:J_bar_sf})
		for $\Delta\epsilon = 4 t_{\text{H}}$
		and the amplitudes of the rapid oscillation are damped with $\mathcal{O}(1/t)$.
		Moreover, expression (\ref{eq:T_min_qm}) for the period of the rapid oscillation -- which was derived only by comparison
		with the classical model -- is confirmed by equation (\ref{eq:approx_dmu=4_current_from_hungarians}).
						
		Since the latter analysis shows a $\mathcal{O}(1/t)$ damping of the rapid oscillation, the open question at hand is, if
		it is reasonable to use our extrapolation method for a quantum system.
		In fact, a pure square wave described by its Fourier series (\ref{eq:rectangular_curve}) has also a damping of the rapid oscillation of 
		$\mathcal{O}(1/t)$ and highly resembles the current described by (\ref{eq:dmu=4_current_from_hungarians}) or depicted in Fig.~\ref{fig:tb_long_time}
		when choosing an appropriate amplitude and number of harmonics.
		Since the mathematical error (\ref{eq:error_for_LC-Line_fft}) of our method stems from the difference of the analyzed curve to a square wave,
		the extrapolation approach provides an even smaller error than (\ref{eq:error_for_LC-Line_fft}) for a quantum system.

\section{\label{section:spinless-fermion-model}Spinless fermion model}

	In this section we discuss our results for the stationary current and the finite-system period
	in the spinless fermion model.  The non-linear dynamics of this model
	and the closely related spin $\frac{1}{2}$ $XXZ$ chain have been studied 
	in several works using td-DMRG methods~\cite{Schmitteckert04,Gobert05,Langer09}. 
	Nevertheless, the steady-state transport properties of the spinless fermion model
	have not been investigated yet. 
	
	\subsection{Current-voltage characteristics}

		Firstly, we remark that the value of the potential bias is given by the parameter $\Delta\epsilon$ of the Hamiltonian from equation (\ref{eq:H_B_def}),
		and it is not the result of a quantum measurement of an observable. Thus it corresponds to an external field applied to the system,
		and in an interacting system it is not always equal to the actual potential difference 
		which is measured in experiments.\cite{Kawabata07,Kawabata96}
		Moreover, in experiments the sample (quantum wire) must be connected by leads
		to the voltage source and measurement apparatus. This modifies the low-energy, 
		low-temperature behaviour of the sample below some crossover energy scale.\cite{Kane92,Safi95}
		Concerning this matter, the linear conductance of a TLL attached to one-dimensional leads has been
		calculated in several works.\cite{Safi95,Maslov95,Ponomarenko95,Janzen06}
		It has been shown that the conductance of this setup is given by $e^2/h$ (which corresponds to equation (\ref{eq:J_bar_sf}) with $K=1$),
		independently of the Coulomb interaction $V_{\text{H}}$ within the wire, and that this result coincides with experimental outcomes.
		
		Our analysis of on-site particle densities shows that the densities for the first and last site of the
		chain only change after $T^{\text{max}}/4$. This is a major reason why we only take values of the current for times smaller 
		than $T^{\text{max}}/4$ into account, as described in our extrapolation approach.
		To test the overall procedure for computing the stationary current
		we have first calculated the current-voltage characteristic of a non-interacting chain ($V_{\text{H}} = 0$).
		This characteristic is known exactly for the first setup\cite{Antal99,Hunyadi04}:
		$\bar{J} = (e/h)\Delta\epsilon$
		for $|eV| \leq 4t_{\text{H}}$ and $\bar{J} = (e/h)4t_{\text{H}}$ for $|eV| \geq 4t_{\text{H}}$, with $\Delta\epsilon = |eV|$ where $V$ is the voltage bias.
		For the second setup, exact results were calculated numerically using the one-particle equation of motion with a system size of $N=1000$.
		
		We have found that our procedure yields stationary current values which agree with an overall error of less than 5\% with the exact results.
		As a second test we have calculated the current-voltage characteristic of the IRLM. The generic shape
		of this curve is known from a field-theoretical analysis, and highly accurate td-DMRG results
		are available for a quantitative comparison.\cite{Boulat08}
		In that regard, we have found that the results from our method agree
        very well with the numerical outcomes from Ref.~\onlinecite{Boulat08}.
		We nevertheless have to remark that we found the IRLM far easier to simulate than the spinless fermion model,
		since for higher $|V_{\text{H}}|$ the correlations (and thus the needed Schmidt dimension) within the spinless fermion model grow more quickly
		than in the IRLM.
		As already mentioned in section \ref{tebd_method} the parameters used for our simulations can result in larger errors for high voltages
		and Coulomb interactions in the TEBD data for the current as a function of time.
		Additionally, finite-size effects also grow with $|V_{\text{H}}|$ and $\Delta\epsilon$.
		This can lead to less accurate extrapolations for the steady-state current in that regime than for non-interacting fermions
		and the IRLM.

		\begin{figure}[t]
			\begin{center}
				\includegraphics[width=1\hsize]{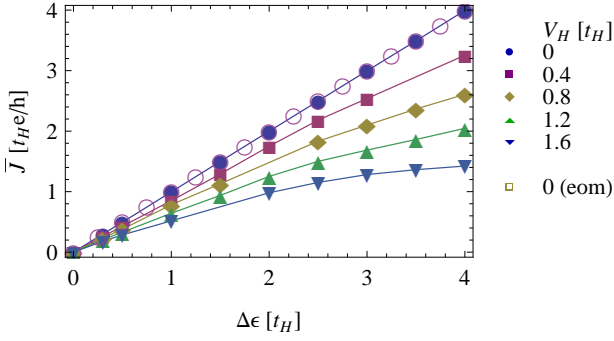}
				\caption{(Color online) Current-voltage curve of the spinless fermion model
						 		 with \textbf{setup (I)} for several \textbf{positive $V_{\text{H}}$}.
						 		 For $V_{\text{H}}=0$ the current has been computed using TEBD and the one-particle equation of motion (eom).
						 		 The lines are guides for the eyes.
				\label{fig:sf_pos_V_setup1}}
			\end{center}
		\end{figure}
		\begin{figure}[t]
			\begin{center}
				\includegraphics[width=1\hsize]{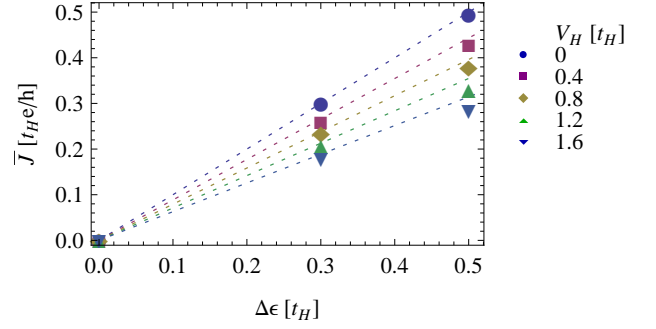}
				\caption{(Color online) Magnification of figure \ref{fig:sf_pos_V_setup1}.
						 		 The lines indicate the linear response according to the TLL theory (\ref{eq:J_bar_sf}) for small $\Delta\epsilon$.
				\label{fig:sf_pos_V_magnification_setup1}}
			\end{center}
		\end{figure}
				
		We now discuss the current-voltage characteristics which have been obtained with the methods
		described above.
		In setup (I) the current of the non-interacting system ($V_{\text{H}}=0$) is strictly proportional 
		to $\Delta \epsilon = |eV|$ up to the band width $4t_{\text{H}}$ and then remains constant.\cite{Antal99,Hunyadi04}
		For $V_{\text{H}} > 0$ we see in Fig.~\ref{fig:sf_pos_V_setup1} that the current increases sub linearly
		with $\Delta \epsilon$ for a fixed interaction strength $V_{\text{H}}$ and that it decreases 
		monotonically with increasing $V_{\text{H}}$ for a fixed potential bias $\Delta \epsilon$.
		The magnification in Fig.~\ref{fig:sf_pos_V_magnification_setup1} shows a comparison
		with the linear response in the TLL theory (\ref{eq:J_bar_sf}) 
		for small $\Delta\epsilon$, using the Bethe Ansatz solution parameters (\ref{eq:TLLparameter}).
		The good agreement confirms the validity of our approach.
		Obviously, an increasing $V_{\text{H}}$ does not only reduce the current but also the 
		range of the potential bias $\Delta\epsilon$ for which the linear response approximation 
		is accurate. 

		\begin{figure}[b]
			\begin{center}
				\includegraphics[width=1\hsize]{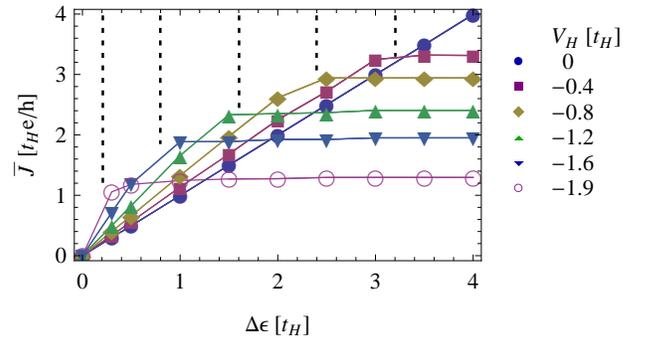}
				\caption{(Color online) Current-voltage curve of the spinless fermion model
						 		 with \textbf{setup (I)} for several \textbf{negative $V_{\text{H}}$}.
						 		 The dashed lines indicate the theoretical beginning of the current plateaus according to (\ref{eq:saturation}).
						 		 The solid lines are guides for the eyes.
				\label{fig:sf_neg_V_setup1}}
			\end{center}
		\end{figure}
		\begin{figure}[t]
			\begin{center}
				\includegraphics[width=1\hsize]{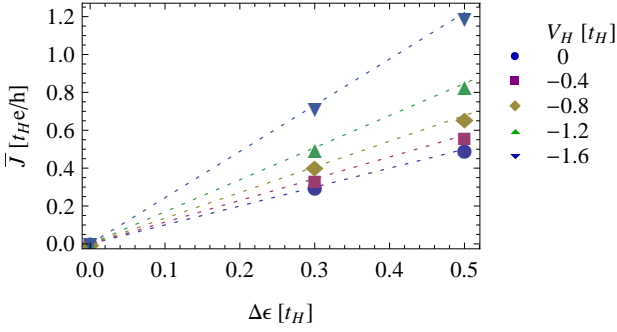}
				\caption{(Color online) Magnification of Fig.~\ref{fig:sf_neg_V_setup1}.
						 		 The lines indicate the linear response according to the TLL theory (\ref{eq:J_bar_sf}) for small $\Delta\epsilon$.
				\label{fig:sf_neg_V_magnification_setup1}}
			\end{center}
		\end{figure}

		In Fig.~\ref{fig:sf_neg_V_setup1} we observe a behaviour for attractive interactions $V_{\text{H}} < 0$
		which is similar to the non-interacting case.
		The current increases almost linearly with $\Delta\epsilon$ up to some $V_{\text{H}}$-dependent threshold 
		potential where it saturates at its maximal value.
		Increasing $|V_{\text{H}}|$ scales the maximal current down but causes a higher linear conductance
		for small $\Delta\epsilon$.
		Figure~\ref{fig:sf_neg_V_magnification_setup1} shows that for negative interactions
		there is also an excellent agreement between our results and the linear response in the
		TLL theory (\ref{eq:J_bar_sf}) for small $\Delta\epsilon$.
		Again increasing $|V_{\text{H}}|$  seems to reduce the
		range of the potential bias $\Delta\epsilon$ for which the linear response approximation 
		is accurate. 

		The occurrence of current plateaus for large potential differences can be easily understood.
		In setup (I) one half of the chain initially contains more particles than the other one due to 
		the applied potential bias $\Delta\epsilon$. When this bias is large enough
		the initial state consists of one completely filled and one completely empty half
		and it does not change for higher $\Delta\epsilon$. In the non-interacting case
		this saturation occurs exactly at $\Delta\epsilon =4t_{\text{H}}$.\cite{Antal99,Hunyadi04}
		When we add an attractive interaction $V_{\text{H}} < 0$ the particles are more likely to stick together and therefore 
		even more particles gather in one half for the same applied potential.
		Thus, saturation (i.e., an initial state with one completely filled and one completely empty half) 
		is reached for smaller $\Delta\epsilon$ as $|V_{\text{H}}|$ increases as seen in Fig.~\ref{fig:sf_neg_V_setup1}.
		We do not understand yet why the plateau heights (i.e. the maximal current) are lowered by increasing $|V_{\text{H}}|$.
		For a repulsive interaction $V_{\text{H}}$ the effect is reversed and less particles gather in one half of the chain
		for a given potential difference when $V_H$ increases. Thus we expect that saturation occurs at higher values 
		$\Delta\epsilon > 4t_{\text{H}}$ beyond the potential range shown in Fig.~\ref{fig:sf_pos_V_setup1}.
		We can easily determine for which potential difference $\Delta\epsilon$
        saturation occurs.
		Removing a single particle from the completely filled reservoir or
		adding one particle to the empty half costs an energy $\Delta\varepsilon / 2 - V_{\text{H}} - 2t_H$.
		Thus saturation occurs if 
		\begin{equation}\label{eq:saturation}
			\Delta\varepsilon \ge 2V_{\text{H}} + 4t_H .
		\end{equation}
		Figure~\ref{fig:sf_neg_V_setup1} shows that this approximation fits well to the numercial data for $V_{\text{H}} \le 0$ whereas
		the saturated regime according to (\ref{eq:saturation}) for
        $V_{\text{H}} > 0$ lies outside the potential range shown in Fig.~\ref{fig:sf_pos_V_setup1},
		as mentioned above.
		
		\begin{figure}[t]
			\begin{center}
				\includegraphics[width=1\hsize]{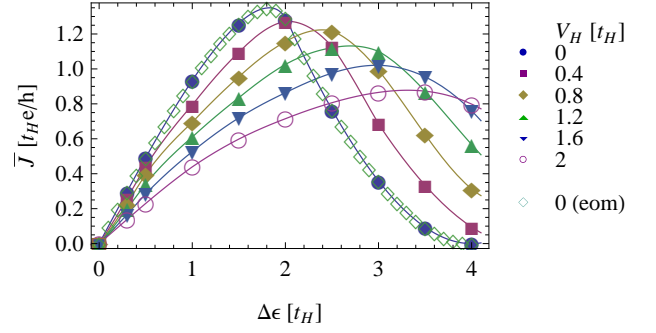}
				\caption{(Color online) Current-voltage curve of the spinless fermion model
						 		 with \textbf{setup (II)} for several \textbf{positive $V_{\text{H}}$}.
						 		 For $V_{\text{H}}=0$ the current has been computed using TEBD and the one-particle equation of motion (eom).
						 		 The lines are guides for the eyes.
				\label{fig:sf_pos_V_setup2}}
			\end{center}
		\end{figure}
		\begin{figure}[b]
			\begin{center}
				\includegraphics[width=1\hsize]{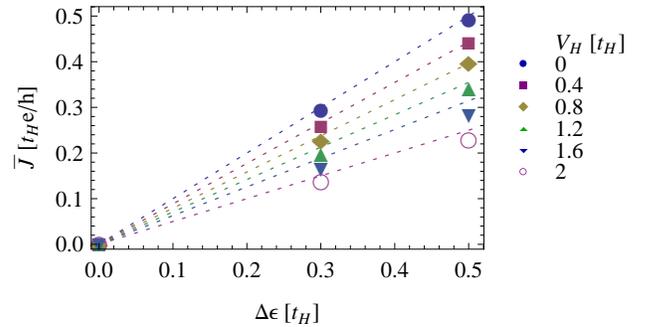}
				\caption{(Color online) Magnification of Fig.~\ref{fig:sf_pos_V_setup2}.
						 		 The lines indicate the linear response according to the TLL theory (\ref{eq:J_bar_sf}) for small $\Delta\epsilon$.
				\label{fig:sf_pos_V_magnification_setup2}}
			\end{center}
		\end{figure}
		
		While for setup (I) the differential conductance is always positive (or zero after saturation), 
		in setup (II) we observe a negative differential conductance.
		Figure~\ref{fig:sf_pos_V_setup2} shows our results for setup (II) with $V_{\text{H}} \geq 0$. 
		The current seems to vanish for very large potential biases
		as predicted by strong-coupling perturbation theory for this setup. 
		We note that the current becomes
		systematically weaker with increasing $V_{\text{H}}$ for a fixed small potential bias.
		For larger $\Delta\epsilon$ the behaviour of $I$ as a function of $V_{\text{H}}$
		is no longer monotonic. 
		In addition, we see that the onset of the
		negative differential conductance (i.e., the position of the maximum of $I$ as a function of $\Delta\epsilon$)
		shifts to higher values with increasing $V_{\text{H}}$.
		The magnification for small $\Delta\epsilon$, Fig.~\ref{fig:sf_pos_V_magnification_setup2},
		confirms again that our results agree with the TLL theory in the linear regime and that
		the range of the linear response regime shrinks with increasing interaction strength.
		
		\begin{figure}[t]
			\begin{center}
				\includegraphics[width=1\hsize]{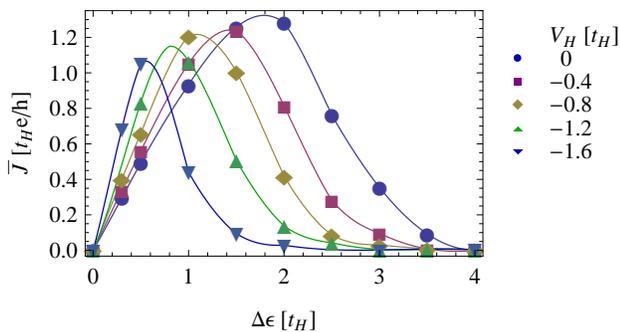}
				\caption{(Color online) Current-voltage curve of the spinless fermion model
						 		 with \textbf{setup (II)} for several \textbf{negative $V_{\text{H}}$}.
						 		 The lines are guides for the eyes.
				\label{fig:sf_neg_V_setup2}}
			\end{center}
		\end{figure}
		\begin{figure}[b]
			\begin{center}
				\includegraphics[width=1\hsize]{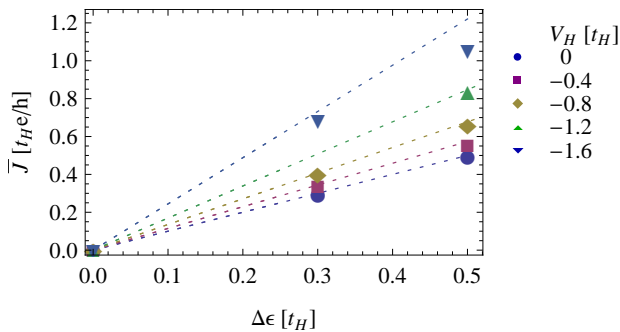}
				\caption{(Color online) Magnification of Fig.~\ref{fig:sf_neg_V_setup2}.
						 		 The lines indicate the linear response according to the TLL theory (\ref{eq:J_bar_sf}) for small $\Delta\epsilon$.
				\label{fig:sf_neg_V_magnification_setup2}}
			\end{center}
		\end{figure}
		
		Results for attractive interactions $V_{\text{H}} <0$ are shown in Fig.~\ref{fig:sf_neg_V_setup2}.
		There we clearly observe that the current vanishes for large potential differences
		as predicted by strong-coupling perturbation theory for setup (II).
		We also see that the current becomes systematically larger with increasing 
		interaction strength $|V_{\text{H}}|$ for a fixed small potential bias.
		For intermediate $\Delta\epsilon$ the behaviour of $I$ as a function of $V_{\text{H}}$
		is not monotonic and for large enough $\Delta\epsilon \gtrsim 2t_{\text{H}}$ the current decreases
		systematically with increasing interaction strength. 
		Moreover, we see that the onset of the negative differential conductance occurs at lower values
		of $\Delta\epsilon$ for increasing $|V_{\text{H}}|$.
		Finally, Fig.~\ref{fig:sf_neg_V_magnification_setup2} shows that in this case
		our results also agree with the TLL theory in the linear regime, but it seems
		that the interaction reduces rapidly the range of potential biases for which the 
		linear response is accurate. 

		The negative differential conductance which is observed in setup (II) for all $V_{\text{H}}$ 
		can be understood in terms of the overlap of the energy bands for elementary excitations.
		Our simulations show that there is no significant difference in the current curve whether one
		initially does or does not couple source and drain (i.e., the two halves of the system). 
		For the following considerations we can therefore regard source and drain
		as initially decoupled, having two separate but equal energy bands.
		Then we apply a step-like voltage in setup (II) and thus shift the two bands against each other.
		For $V_{\text{H}}=0$ only overlapping occupied one-particle states in the source and unoccupied ones in the drain 
		contribute to the current.
		While this overlap rises with increasing $|\Delta\epsilon|$ from $0$ to $2t_{\text{H}}$, it diminishes from 
		$|\Delta\epsilon| = 2t_{\text{H}}$ and reaches zero at $4t_{\text{H}}$.
		As a result, the current is maximal for $\Delta\epsilon \approx 2t_{\text{H}}$ and vanishes for large $\Delta\epsilon$.
		Similarly, we can understand the non-monotonic behaviour in interacting cases ($V_{\text{H}} \neq 0$) 
		in terms of the overlap of the elementary excitation bands in the spinless Luttinger liquids
		in the two halves of the system.
		However, the effective bandwidth is renormalized exactly as the Fermi velocity in equation
		(\ref{eq:velocity}).
		Therefore, the maximum of the current is reached for a smaller potential bias $\Delta\epsilon$ when $V_{\text{H}}$
		becomes negative as shown in Fig.~\ref{fig:sf_neg_V_setup2},
		and shifted to a higher potential bias when $V_{\text{H}}$ increases as seen in Fig.~\ref{fig:sf_pos_V_setup2}.
		Our conclusion agrees with the findings in Ref.~\onlinecite{Baldea10} where it is shown that within a similar one-dimensional
		model a negative differential conductance is mainly caused by finite electrode bandwidths.

	\subsection{Influence of $V_{\text{H}}$ on the period $T^{\text{max}}$ in the finite quantum system}
		
		Our results show a further effect of $V_{\text{H}}$ on the finite-system current. While for $V_{\text{H}} > 0$ the period of the rectangular oscillation
		becomes smaller, it grows for negative interactions.
		
		\begin{figure}[t]
			\begin{center}
				\includegraphics[width=1\hsize]{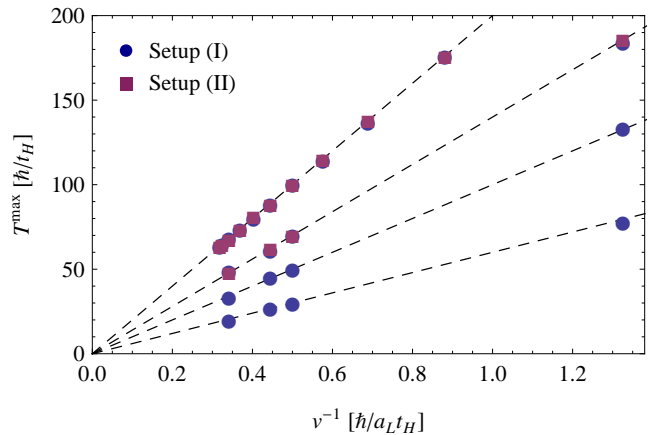}
				\caption{(Color online) Period $T^{\text{max}}$ of the rectangular current oscillation in the finite spinless fermion model
				 				 as a function of the inverse renormalized Fermi velocity $v^{-1}$ from expression (\ref{eq:velocity}).
				 				 The dashed lines are described by equation (\ref{eq:T_max_vs_VH}).
				\label{fig:sf_periods}}
			\end{center}
		\end{figure}
		
		Figure~\ref{fig:sf_periods} shows the period $T^{\text{max}}$ of the rectangular current oscillation in the finite system
		as a function of the inverse renormalized Fermi velocity $v^{-1}$ from expression (\ref{eq:velocity}).
		The $T^{\text{max}}$ values are mean values from various $\Delta\epsilon$ for each setup.
		These values agree perfectly with the dashed lines in Fig.~\ref{fig:sf_periods}  
		which are given by the equation
		\begin{equation}\label{eq:T_max_vs_VH}
			T^{\text{max}}(V_{\text{H}}) = \frac{v(V_{\text{H}}=0)}{v(V_{\text{H}})} T^{\text{max}}(V_{\text{H}}=0) .
		\end{equation}
		Thus, the period is fully determined by the time-scale of the non-interacting system 
		$T^{\text{max}}(V_{\text{H}}=0)$ from (\ref{eq:T_qm}) and 
		the renormalized Fermi velocity $v$ from expression (\ref{eq:velocity}). 
		
\section{Summary and conclusion}
	
	We have presented a method for investigating the non-linear steady-state transport properties of
	one-dimensional correlated quantum systems. 
	First, we have analyzed finite-size effects and transient currents in a related classical model,
	the so-called $LC$ line.
	This exact analysis shows how steady-state currents in the thermodynamic limit can be calculated
	from transient currents in finite systems.
	We have found that for currents in quantum systems, in particular the spinless fermion model, finite-size effects
	can be understood from the behaviour of the $LC$ line.
	
	On the basis of the strong connection between the finite-size behaviour in classical and quantum models
	we have developed a method to extract a stationary current from
	simulations of finite-size quantum systems. We only need numerical data $I(t)$ up to a time of the order of $T^{\text{max}}/4$ where
	$T^{\text{max}}\propto N$ is the period of the approximately rectangular signal $I(t)$ in a quantum system of size $N$.
	The non-equilibrium dynamics of correlated quantum systems has been calculated using the 
	time-evolving block decimation method (TEBD).
	We have implemented a multi-threaded version of TEBD with an extremely low overhead (less than 1\% for 10 threads) which
	allows us to simulate the non-equilibrium dynamics up to the time-scale $2T^{\text{max}}$.
	
	We have determined the full $I$--$V$ characteristic of the spinless fermion model with 
	nearest-neighbour hopping $t_{\text{H}}$ and interaction $V_{\text{H}}$ using two different setups to generate currents.
	In setup (I) the initial state has different particle numbers in its two halves due to an applied
	potential difference while the system evolves in time with an overall equal on-site potential.
	In setup (II) we calculate the initial state without a potential difference but turn it on for the real time evolution.
	For non-interacting fermions ($V_{\text{H}}=0$) our outcomes agree with the analytical solutions
	and with the results of the one-particle equation of motion method.
	Additionally, we have checked that our results coincide with the field-theoretical analysis
	and td-DMRG simulations\cite{Boulat08} for the IRLM.
	
	For interacting fermions we have found that the steady-state current is independent from the setup in the linear regime $|eV| \ll 4t_{\text{H}}$.
	Moreover, our numerical data agree with the predictions of the Luttinger liquid theory combined with the Bethe Ansatz solution. 
	For larger potentials $V$ the steady-state current depends on the current-generating setup. 
	This difference is well understood for non-interacting systems~\cite{Branschaedel10,Baldea10}
	but  can only be explained qualitatively for our interacting system.
	With setup (I) we have found that the current increases with $V$ and saturates at a finite value 
	when the potential difference is high enough to separate the initial state in one filled and one empty half.
	With setup (II) we observe a negative differential conductance that can be understood in terms of the overlap of the 
	elementary excitation bands in the spinless Luttinger liquids in the two halves of the system.
	Both effects -- the plateaus and the negative differential conductance at large potential bias 
	-- are due to the finite bandwidth of the system
	which agrees with the findings in Ref.~\onlinecite{Baldea10}.
	However, in our case the interaction renormalizes the effective bandwidth exactly as the Fermi velocity in equation (\ref{eq:velocity})
	so that the current maxima and the onset of the plateaus depend on the strength of the interaction $V_{\text{H}}$.
	The change in the typical time-scale $T^{\text{max}}$ also solely depends on the renormalized Fermi velocity.
	
	The methods presented in this work can be applied to other systems, such as models of quantum dots or wires coupled to leads including
	electronic and also bosonic degrees of freedom.
	We have already checked the applicability of these methods to the IRLM, the one-dimensional Hubbard model away from half-filling
	and also two-leg ladder systems.
	Hence, we believe that they will be very useful to study non-linear transport properties of correlated low-dimensional conductors.
    Although we have mostly tested our appraoch in the limit of transparent
    coupling between source and drain, we think that it can also be applied when
    the hybridization between leads is very small.  This issue will be examined in future works.

\begin{acknowledgments}

	We thank P. Schmitteckert for useful discussions. 
	This work is supported by the	\textit{NTH school for contacts in nanosystems}.
	Computational resources for this work were provided by the Regional Computing Center for Lower Saxony at the 
	Leibniz Universit\"at Hannover (RRZN).

\end{acknowledgments}

\appendix

\section{Approximations of the Bessel und Struve functions}

		The Bessel functions of the first kind and Struve functions have the following approximations which become exact for $x\rightarrow\infty$
		\begin{eqnarray}\label{eq:approx_bessel_struve}
				\BesselJ_0(x) &=& \sqrt{\frac{2}{\pi x}} \cos\left( x - \frac{\pi}{4} \right) + \mathcal{O}\left(\frac{1}{x}\right) ,\nonumber\\
				\BesselJ_1(x) &=& \sqrt{\frac{2}{\pi x}} \cos\left( x - \frac{3}{4}\pi \right) + \mathcal{O}\left(\frac{1}{x}\right) ,\nonumber\\						
				\StruveH_0(x) &=& \sqrt{\frac{2}{\pi x}} \sin\left( x - \frac{\pi}{4} \right) + \mathcal{O}\left(\frac{1}{x}\right) ,\nonumber\\
				\StruveH_1(x) &=& \frac{2}{\pi} - \sqrt{\frac{2}{\pi x}}\cos\left( x-\frac{\pi}{4} \right) + \mathcal{O}\left(\frac{1}{x}\right) .
		\end{eqnarray}
		Using these expressions together with equation (\ref{eq:i(t)}) yields the stationary current value (\ref{eq:infinite_current})
		but does not provide an approximate expression for the short-time behaviour of the current, since all
        time-dependent terms cancel out.
		Instead, the asymptotic series expansions from
        Refs.~\onlinecite{Abramowitz72} and~\onlinecite{Erdelyi56}
		and the approximation of $\StruveH_1(x)$\cite{Aarts03}
		\begin{eqnarray}\label{eq:asymptotic_approximations}
				\BesselJ_0(x) &=& \frac{(8x-1)\cos(x)+(8x+1)\sin(x)}{8\sqrt{\pi x^3}} + \mathcal{O}\left(\frac{1}{x^{\frac{5}{2}}}\right) ,\nonumber\\
				\BesselJ_1(x) &=& \frac{(3-8x)\cos(x)+(3+8x)\sin(x)}{8\sqrt{\pi x^3}} + \mathcal{O}\left(\frac{1}{x^{\frac{5}{2}}}\right) ,\nonumber\\						
				\StruveH_0(x) &=& \frac{8x\sqrt{\pi}(\sin(x)-\cos(x)) + 16\sqrt{x} - \sqrt{\pi}}{8\pi\sqrt{x^3}} \nonumber\\
				              & & \quad +~ \mathcal{O}\left(\frac{1}{x^{\frac{5}{2}}}\right) ,\nonumber\\
				\StruveH_1(x) &\approx& \frac{2}{\pi}+\frac{\left(\frac{16}{\pi}-5\right)\sin{x}}{x} \nonumber\\
				              & & \quad +~ \frac{\left(12-\frac{36}{\pi}\right)(1-\cos(x))}{x^2}	- \BesselJ_0(x)
		\end{eqnarray}
		result in the expression (\ref{eq:I_gg_final}).

\section{\label{app:period}Period of the rectangular oscillation in the tight-binding model}

		The time evolution of the single particle reduced density matrix (\ref{eq:time_evolution_Greens_function})
		is given by
		\begin{equation}
			\mathcal{G}_{ij}(t) = \exp{\left ( - \frac{i}{\hbar} \varepsilon_{q} t \right )} \mathcal{G}_{kq}(0) \exp{\left ( \frac{i}{\hbar} 
																		\varepsilon_{k} t \right )}
		\end{equation}
		in the eigenbasis of a time-constant single particle Hamiltonian $H^{(1)}$.
		$\Psi_{k,i}$ denotes the $i$-th component of the $k$-th eigenvector of $H^{(1)}$ and $\varepsilon_{k}$
		the corresponding eigenvalue. A Fourier transformation gives
		\begin{equation}
			\tilde{\mathcal{G}}_{kq}(\omega) = \delta \left [\varepsilon_{k} - \varepsilon_{q} - \hbar\omega \right ] \mathcal{G}_{kq}(0) .
		\end{equation}
		For the tight-binding model with zero on-site potentials one has
		\begin{eqnarray}\label{eq:eigenvector_of_H}
		 		\Psi_{k,i} &=& \sqrt{\frac{2}{N+1}} \sin\left(\frac{k i \pi}{N+1}\right) ,\nonumber\\
		 		\varepsilon_{k} &=& - 2t_{\text{H}}\cos\left( \frac{k\pi}{N+1} \right) .
		\end{eqnarray}		    
		The period of the largest (rectangular) oscillation is then given by
		\begin{eqnarray}\label{eq:T_qm}
			T^{\text{max}} = \frac{\pi\hbar}{t_{\text{H}} \sin\left(\frac{\pi}{N+1}\right)} \approx \hbar \cdot \frac{N}{t_{\text{H}}} \quad \text{for} \quad N \gg 1.
		\end{eqnarray}

\section{\label{app:current_in_quantum_system}Current in the infinite one-dimensional tight-binding model}

		The expectation value of the current for setup (I) and for initial conditions like in Fig.~\ref{fig:tb_long_time} (one completely filled and one
		completely empty half) is for $N\rightarrow\infty$ given by\cite{Antal99,Hunyadi04}
		\begin{equation}\label{eq:hungarian_current}
			J_{\text{F}}^k(t) = \frac{2et_{\text{H}}}{\hbar} \sum_{l=k-\frac{N}{2}}^{\infty} \BesselJ_l(\omega t) \BesselJ_{l+1}(\omega t)
		\end{equation}
		where $k$ denotes the site, $\omega = 2 t_{\text{H}}/\hbar$ and $\BesselJ_l(z)$ are the Bessel functions of the first kind.
		Reformulating the sum and utilizing the Bessel recursion relation
		\begin{equation}
			\frac{2l}{z} \BesselJ_l(z) = \BesselJ_{l+1}(z) + \BesselJ_{l-1}(z)
		\end{equation}
		gives for $z \ne 0$ and $k = N/2$
		\begin{equation}
				J_{\text{F}}(\omega t) = J_{\text{F}}^{N/2}(\omega t) = \frac{4et_{\text{H}}}{\hbar \omega t} 
																 \sum_{l=0}^{\infty} \left[ (\BesselJ_{2l+1}(\omega t))^2 \cdot (2l+1) \right] .
		\end{equation}
		With\cite{Abramowitz72}
		\begin{eqnarray}
				& &\sum_{l=0}^{\infty} (4l+2\nu+2) \BesselJ_{2l+\nu+1}(z) \BesselJ_{2l+\nu+1}(w)\nonumber\\
				& &= \frac{zw}{z^2-w^2}\left[ z\BesselJ_{\nu+1}(z)\BesselJ_{\nu}(w)-w\BesselJ_{\nu}(z)\BesselJ_{\nu+1}(w) \right]
		\end{eqnarray}
		for $\text{Re}(\nu) > -1$ and $z \ne w$ the current is given by
		\begin{equation}
				J_{\text{F}}(t) = \frac{2 e t_{\text{H}}}{\hbar} \lim_{t' \rightarrow t} 
						\frac{\omega t\BesselJ_1(\omega t)\BesselJ_0(\omega t')-\omega t'\BesselJ_0(\omega t)\BesselJ_1(\omega t')}
						     {(\pi \omega t')^{-1}((\omega t)^2 - (\omega t')^2)} .
		\end{equation}
		Applying l'H\^ospital's rule one gets
		\begin{equation}\label{eq:dmu=4_current_from_hungarians_app}
				J_{\text{F}}(t) = \frac{et_{\text{H}}}{\hbar} \omega t \left[ (\BesselJ_0(\omega t))^2 + (\BesselJ_1(\omega t))^2 \right] .
		\end{equation}

\end{document}